\documentclass[fleqn,usenatbib]{mnras}

\usepackage{newtxtext,newtxmath}

\usepackage[T1]{fontenc}

\DeclareRobustCommand{\VAN}[3]{#2}
\let\VANthebibliography\thebibliography
\def\thebibliography{\DeclareRobustCommand{\VAN}[3]{##3}\VANthebibliography}

\usepackage{graphicx}	
\usepackage{amsmath}	
\usepackage{multirow}

\newcommand{\Mg}{M_\mathrm{g}}
\newcommand{\tdep}{t_\mathrm{dep}}
\newcommand{\epsn}{\epsilon_{p,\mathrm{SN}}}

\newcommand{\epw}{\epsilon_{p,\mathrm{w}}}

\newcommand{\depw}{\delta_{\mathrm{w}}}

\newcommand{\sip}{\sigma_{\mathrm{ion},p}}
\newcommand{\sie}{\sigma_{\mathrm{ion},e}}
\newcommand{\Lp}{\mathcal{L}_p}
\newcommand{\Lpion}{\mathcal{L}_{\mathrm{ion},p}}
\newcommand{\Lpinel}{\mathcal{L}_{\mathrm{inel},p}}
\newcommand{\Le}{\mathcal{L}_e}
\newcommand{\Leion}{\mathcal{L}_{\mathrm{ion},e}}
\newcommand{\Leionone}{\mathcal{L}_{\mathrm{ion},1,e}}
\newcommand{\Lebrem}{\mathcal{L}_{\mathrm{brem},e}}

\newcommand{\nH}{n_\mathrm{H}}
\newcommand{\Psiip}{\Psi_{\mathrm{ion},p}}
\newcommand{\Psiep}{\Psi_{\gamma,p}}
\newcommand{\Psiie}{\Psi_{\mathrm{ion},e}}
\newcommand{\Psiee}{\Psi_{\gamma,e}}
\newcommand{\Psiipavg}{\langle\Psi_{\mathrm{ion},p}\rangle}
\newcommand{\Psiepavg}{\langle\Psi_{\gamma,p}\rangle}
\newcommand{\Psiieavg}{\langle\Psi_{\mathrm{ion},e}\rangle}
\newcommand{\Psieeavg}{\langle\Psi_{\gamma,e}\rangle}
\newcommand{\Eg}{E_\gamma}
\newcommand{\Tip}{T_{\mathrm{i},p}}
\newcommand{\Tie}{T_{\mathrm{i},e}}

\newcommand{\add}[1]{{#1}}

\newcommand{\aref}[1]{\hyperref[#1]{Appendix~\ref{#1}}}

\newcommand{\noopsort}[2]{#2}

\usepackage{scalerel,tikz}
\usetikzlibrary{svg.path}
\definecolor{orcidlogocol}{HTML}{A6CE39}
\tikzset{orcidlogo/.pic={
 \fill[orcidlogocol] svg{M256,128c0,70.7-57.3,128-128,128C57.3,256,0,198.7,0,128C0,57.3,57.3,0,128,0C198.7,0,256,57.3,256,128z};
 \fill[white] svg{M86.3,186.2H70.9V79.1h15.4v48.4V186.2z}
 svg{M108.9,79.1h41.6c39.6,0,57,28.3,57,53.6c0,27.5-21.5,53.6-56.8,53.6h-41.8V79.1z M124.3,172.4h24.5c34.9,0,42.9-26.5,42.9-39.7c0-21.5-13.7-39.7-43.7-39.7h-23.7V172.4z}
 svg{M88.7,56.8c0,5.5-4.5,10.1-10.1,10.1c-5.6,0-10.1-4.6-10.1-10.1c0-5.6,4.5-10.1,10.1-10.1C84.2,46.7,88.7,51.3,88.7,56.8z};
}}
\newcommand\orcidicon[1]{\href{https://orcid.org/#1}{\mbox{\scalerel*{
\begin{tikzpicture}[yscale=-1,transform shape]
\pic{orcidlogo};
\end{tikzpicture}
}{|}}}}

\title[Cosmic ray budgets]{The cosmic ray ionisation and $\gamma$-ray budgets of star-forming galaxies}

\author[M. R. Krumholz et al.]{Mark R. Krumholz$^{\orcidicon{0000-0003-3893-854X}1, 2}$\thanks{E-mail: mark.krumholz@anu.edu.au (MRK)},
Roland M. Crocker$^{\orcidicon{0000-0002-2036-2426}1}$
and
Stella S. R. Offner$^{\orcidicon{0000-0003-1252-9916}3}$
\\
$^{1}$Research School of Astronomy and Astrophysics, Australian National University, Canberra ACT 2600, Australia\\
$^{2}$ARC Center of Excellence for Astronomy in Three Dimensions (ASTRO-3D), Canberra ACT 2600, Australia\\
$^{3}$Astronomy Department, University of Texas, Austin TX 78712, USA
}

\date{Accepted XXX. Received YYY; in original form ZZZ}

\pubyear{2023}

\begin{document}
\label{firstpage}
\pagerange{\pageref{firstpage}--\pageref{lastpage}}
\maketitle

\begin{abstract}
Cosmic rays in star-forming galaxies are a dominant source of both diffuse $\gamma$-ray emission and ionisation in gas too deeply shielded for photons to penetrate. Though the cosmic rays responsible for $\gamma$-rays and ionisation are of different energies, they are produced by the same star formation-driven sources, and thus galaxies' star formation rates, $\gamma$-ray luminosities, and ionisation rates should all be linked. In this paper we use up-to-date cross-section data to determine this relationship, finding that cosmic rays in a galaxy of star formation rate $\dot{M}_*$ and gas depletion time $\tdep$ produce a maximum primary ionisation rate $\zeta \approx 1\times 10^{-16} (\tdep/\mbox{Gyr})^{-1}$ s$^{-1}$ and a maximum $\gamma$-ray luminosity $L_\gamma\approx 4\times 10^{39} (\dot{M}_*/\mathrm{M}_\odot\mbox{ yr}^{-1})$ erg s$^{-1}$ in the 0.1 - 100 GeV band. These budgets imply either that the ionisation rates measured in Milky Way molecular clouds include a significant contribution from local sources that elevate them above the Galactic mean, or that CR-driven ionisation in the Milky Way is enhanced by sources not linked directly to star formation. Our results also imply that ionisation rates in starburst systems are only moderately enhanced compared to those in the Milky Way. Finally, we point out that measurements of $\gamma$-ray luminosities can be used to place constraints on galactic ionisation budgets in starburst galaxies that are nearly free of systematic uncertainties on the details of cosmic ray acceleration.
\end{abstract}




\section{Introduction}
\label{sec:intro}

Cosmic rays (CR), the non-thermal particles accelerated by interstellar shocks, play an important role in multiple distinct areas of astrophysics. In galaxy formation theory, study of CRs as a potential source of feedback capable of regulating star formation and driving galactic winds has undergone a renaissance in the last decade \citep[e.g.,][]{Socrates08a, Uhlig12a, Salem14a, Girichidis18a, Chan19a, Hopkins20a, Yu20a, Crocker21a, Crocker21b}. In $\gamma$-ray\add{, neutrino,} and radio astronomy, high-energy CRs are the dominant sources of emission from star-forming galaxies at both long wavelengths \citep[e.g.,][]{Condon92a, Brown17a} and at photon energies $\gtrsim 0.1$ GeV \add{and neutrino energies $\gtrsim 1$ TeV} \citep[e.g.,][]{Yoast-Hull16a, Peretti19a, Roth21a, Ha21a}. In astrochemistry, low-energy CRs are dominant drivers of both heating and chemistry in dense gas that is shielded from interstellar radiation fields (e.g., \citealt{Glassgold12a, Padovani15a, Gaches18a}; see \citealt{Padovani20a} and \citealt{Gabici22a} for recent reviews).

CRs are thought to be accelerated by interstellar shocks, with shocks driven by SNe as the dominant contributor averaged over galactic scales \citep{Caprioli12a, Bell13a}. Since core collapse SNe rapidly follow star formation, it is therefore natural to expect a linear relationship between star formation rate and CR injection into a galaxy, and thus at least potentially between star formation rate and non-thermal emission that traces CRs. The extent to which such a relationship holds, and to which particular galaxies deviate from it, can then be interpreted as constraining the fraction of CRs that escape from galaxies; this in turn can be used to illuminate the physics of CR transport through interstellar gas \citep[e.g.,][]{Lacki10a, Lacki10b, Lacki11a, Krumholz20a, Ajello20a, Kornecki20a, Kornecki22a, Crocker21a, Werhahn21b, Werhahn21c, Ambrosone22a, Owen22a, Peretti22a}. A crucial input to these interpretive efforts is the total $\gamma$-ray production budget associated with star formation -- i.e., in a galaxy that is perfectly calorimetric, such that all the CRs accelerated by young stars and their feedback give up their energy within the galaxy, what $\gamma$-ray luminosity would we expect per unit mass of stars formed? A number of authors have attempted to compute this number \citep[e.g.,][]{Lacki11a, Kornecki20a, Crocker21a, Werhahn21a}, but the inputs to these calculations often do not represent the state of the art in either particle physics or modeling of star formation; for example, none of the papers cited attempts to estimate the contribution to $\gamma$-ray emission from CR sources other than SNe (likely subdominant, but perhaps not completely negligible), none take into account the most recent results from the SN community about which stars are likely to end their lives as SNe \citep[e.g,][]{Sukhbold16a}, and all but \add{a few of the most recent} compute $\gamma$-ray emission using models for pionic $\gamma$-ray production that precede the launch of \textit{Fermi} \citep[e.g.,][]{Kelner06a} and that have proven to be substantially inaccurate at $\gamma$-ray energies $\lesssim 1$ GeV. One of our goals in this paper it to provide a calibration of the $\gamma$-ray production budget associated with star formation that improves on earlier calibrations by remedying these issues.

While the $\gamma$-ray budget of star formation has received considerable attention, the ionisation budget has not, despite the underlying question being quite similar: given a certain star formation rate, and thus a certain rate at which CRs are accelerated, for a fully calorimetric galaxy what ionisation rate would we expect those CRs to be able to produce in dense, shielded gas where CRs are the only significant ionisation source? Put another way, what is the CR ionisation budget due to star formation? Providing a first calculation of this number, and its relationship to the $\gamma$-ray production budget, is the second goal of this paper.

The question of the ionisation budget is particularly urgent due to recent interest, both observational and theoretical, in the ionisation rate and chemical state of starburst galaxies. On the theoretical side, a number of authors have investigated how the chemistry of molecular gas changes when it is subjected to ionisation rates far beyond those found in the Milky Way, as might be expected in galaxies undergoing much more intense star formation \citep[e.g.,][]{Papadopoulos10a, Meijerink11a, Bisbas15a, Bisbas17a, Bialy15a, Narayanan17a, Papadopoulos18a, Krumholz20a}. However, the exact chemical state depends sensitively on how extreme the ionisation rate is compared to the $\approx 10^{-16}$ s$^{-1}$ typical of Milky Way molecular clouds \citep[e.g.,][]{Indriolo12a, Indriolo15a}. For example, ionisation rates enhanced by factors of $\lesssim 100$ compared the Milky Way still yield CO as the dominant chemical state of carbon in dense, UV-shielded gas, while higher ionisation rates lead to atomic C as the dominant species \citep[e.g.,][]{Bisbas15a}. In the absence of theoretical guidance, it is difficult to know which of these is a more realistic prospect. Different plausible assumptions -- e.g., that the ionisation rate is proportional to the total star formation rate versus the star formation rate per unit area versus the star formation rate per unit volume -- lead to very different conclusions.

Observationally, studies of starburst galaxies in both the local Universe \citep[e.g.,][]{Gonzalez-Alfonso13a, Gonzalez-Alfonso18a, van-der-Tak16a} and at high redshift \citep[e.g.,][]{Muller16a, Indriolo18a, Kosenko21a} report an immense range of values, from those only mildly enhanced relative to the Milky Way to those that are $\sim 5-6$ orders of magnitude larger. At least part of this range likely reflects the fact that there is no single ionisation rate in such galaxies: many starbursts contain active galactic nuclei (AGN) that can drive very high ionisation rates close to the AGN, but this may then coexist with much more modest ionisation rates in the majority of the gas. A spatially unresolved measurement, or an absorption measurement along a pencil beam to a background source, mixes together these regions of different ionisation rate in an unknown and poorly-constrained way. This in turn makes measured ionisation rates very difficult to interpret. Again, we are confronted with a situation where some theoretical guidance on what sorts of ionisation rates are realistic for starbursts would be helpful. 

Given these motivations, the remainder of this paper is organised as follows. In \autoref{sec:efficiencies} we define the efficiency of ionisation and $\gamma$-ray production by CRs, and calculate these efficiencies as a function of CR energy for both protons and electrons. In \autoref{sec:budgets}, we use our calculated efficiencies to estimate the ionisation and $\gamma$-ray budgets of star-forming galaxies as a function of their properties. We discuss the implications of our findings for both the Milky Way and extragalactic systems in \autoref{sec:discussion}, and then we summarise our findings and discuss future prospects in \autoref{sec:conclusions}.

\section{Ionisation and $\gamma$-ray production efficiencies}
\label{sec:efficiencies}

Our goal in this section is to determine how efficiently cosmic rays that are injected into the interstellar gas in a galaxy can be converted into ionisations and observable $\gamma$-ray emission. We will ultimately derive our final results for these quantities from numerical Monte Carlo calculations of CR evolution using the \textsc{criptic} CR propagation code \citep{Krumholz22a}. However, before beginning the numerical calculations, it is of benefit to develop a simple analytic model using the continuous slowing-down approximation (\citealt{Fano53a}; \autoref{ssec:csda}), whereby we approximate loss of energy by CRs as a continuous process. This treatment provides insight that will be helpful to keep in mind when exploring the numerical results. We then proceed to those full numerical results in \autoref{ssec:criptic_numerical}, and use these results to derive spectral-averaged CR ionisation and $\gamma$-ray production efficiencies in \autoref{ssec:spectral_average}.

\subsection{The continuous slowing down approximation}
\label{ssec:csda}

We begin by considering the fate of a single CR of initial kinetic energy $T_i$ that is injected into a galaxy, and that continues to interact with interstellar material until it looses all its energy and again becomes part of the thermal population. Our first approach to this problem is to use the continuous slowing down approximation (CSDA) whereby we approximate processes that cause large, discontinuous jumps in CR energy (e.g., a pion-producing collision between a CR proton and an ISM proton) as instead causing continuous energy loss at a rate that matches the average loss rate caused by the discontinuous jumps.

\subsubsection{Protons}
\label{sssec:proton_ion}

Let $\sip$ be the ionisation cross section for collisions between the CR and a background gas\footnote{Note here that we are counting only primary ionisations caused by the proton itself, not secondary ionisations causes when the low-energy electrons produced by the primary ionisations collide with other neutral atoms or molecules. \add{We do not include secondary ionisations because the convention in the astrochemistry literature is to report the inferred primary ionisation rate, so this is the quantity we want to compute. The ionisation cross section including secondary ionisations would be a factor of $\approx 1.7 - 2$ larger, depending on the chemical state of the background gas and the proton energy \citep{Ivlev21a}.}}, and let $d\sigma_{\gamma,p}/d\Eg$ be the differential cross section for inelastic nuclear interactions leading to production of $\gamma$-ray photons with energy $\Eg$, summing over all possible production channels for which the final state particles include photons; the dominant channel is generally $pp\to pp\pi^0\to pp2\gamma$. We define these cross sections to be measured per H nucleus in the background gas, so for a background gas with number density of H nuclei $\nH$, the proton therefore causes ionisations and produces photons with energy from $\Eg$ to $\Eg + d\Eg$ at a rate per unit time $\dot{N}_{\rm ion} = \nH\sip \beta c$ and $d\dot{N}_\gamma/d\Eg = \nH (d\sigma_{\gamma,p}/d\Eg) \beta c$, respectively, where $\beta$ is the proton velocity normalised to $c$.

In a fully neutral medium, ionisations and nuclear inelastic collisions are the only significant energy loss mechanisms. For the former, we can write the loss rate as
\begin{equation}
    \dot{T}_{\mathrm{ion},p} = \nH \beta c \int_0^{W_{\rm max}} (W + I) \frac{d\sip}{dW} \, dW \equiv \nH \beta c \Lpion,
    \label{eq:ion_loss_p}
\end{equation}
where $d\sip/dW$ is the differential cross section for production of an ejected electron of kinetic energy $W$, $I$ is the ionisation potential of the gas being ionised, $W_{\rm max} = 4 (m_e/m_p) T_p - I$ is the maximum ejected electron kinetic energy allowed by kinematics, and we have implicitly defined the proton loss function $\Lpion$. For the purposes of our CSDA calculation, we approximate energy loss due to inelastic collisions by assuming that each collision removes $\approx 1/2$ of the current proton kinetic energy \citep{Gaisser90a}. Consequently, we can write the inelastic collision loss rate as
\begin{equation}
    \dot{T}_{\mathrm{inel},p} = \nH \beta c \sigma_{\rm inel} \frac{T_p}{2} \equiv \nH \beta c \Lpinel,
\end{equation}
where $\sigma_{\rm inel}$ is the total inelastic collision cross section, and we have defined the inelastic collision loss function in analogy to the ionisation one.

Given these expressions, the number of ionisations per unit change in proton kinetic energy is $dN_{\rm ion}/dT_p = \sip / \Lp$, where $\Lp = \Lpion + \Lpinel$ is the total proton loss function, and the \textit{total} number of ionisations that an injected CR proton with initial energy $\Tip$ is capable of causing is
\begin{equation}
    N_{\mathrm{ion},p} = \int_0^{\Tip} \frac{\sip}{\Lp} \, dT_p
    \label{eq:Nion}
\end{equation}
Performing the analogous procedure for $\gamma$-ray production gives
\begin{equation}
    \frac{dN_{\gamma,p}}{d\Eg} = \int_0^{\Tip} \frac{d\sigma_{\gamma,p}/d\Eg}{\Lp} \, dT_p
    \label{eq:dNdEgamma}
\end{equation}
which is the total number of $\gamma$-ray photons per unit photon energy that a CR proton of initial energy $\Tip$ is capable of producing; integrating this emission over an energy range from $E_0$ to $E_1$, the total $\gamma$-ray luminosity that a CR proton can produce is
\begin{equation}
    E_{\gamma,p}\!\left(E_0,E_1\right) = \int_{E_0}^{E_1} E_\gamma \frac{dN_{\gamma,p}}{dE_\gamma} \, dE_\gamma.
\end{equation}

It is convenient to express these quantities in terms of a dimensionless efficiency. We therefore define the ionisation and $\gamma$-ray production efficiencies as
\begin{eqnarray}
    \Psiip & \equiv & \frac{N_{\mathrm{ion},p} I}{\Tip} \\
    \Psiep\!(E_0,E_1) & \equiv & \frac{E_{\gamma,p}\!\left(E_0,E_1\right)}{\Tip}.
    \label{eq:Psiep}
\end{eqnarray}
These quantities have straightforward physical meanings: $\Psiip$ is the number of ionisations caused compared to the maximum number possible given the CR energy and the ionisation potential of the gas, while $\Psiep$ is the fraction of the initial CR energy that is radiated into $\gamma$-rays with energies in the range $(E_0, E_1)$. We defer numerical evaluation of them to \autoref{ssec:criptic_numerical}.

\subsubsection{Electrons}
\label{sssec:electron_ion}

Developing a CSDA model for electrons is somewhat more complex, because electrons are subject to loss mechanisms -- synchrotron and inverse Compton (IC) radiation -- whose rates are \textit{not} proportional to the number density of the background gas. Consequently, we cannot obtain expressions for ionisation and photon production that are independent of interstellar environment; these quantities will necessarily depend on the importance of synchrotron and IC losses, both relative to each other and relative to the other loss mechanisms that do operate at rates proportional to $\nH$. We therefore parameterise the importance of synchrotron and IC losses as follows: under the assumption that CR electrons are relativistic\footnote{In fact, this assumption is not strictly necessary, since by the time CR electrons reach energies such that they are no longer relativistic, synchrotron and IC losses -- the ones where our expressions depend on this assumptions, are generally unimportant in any event. Nonetheless, we make this assumption explicit to caution readers that our expressions for these two rates do assume that the electrons are at least trans-relativsitic.} and in the Thomson limit for IC scattering, the energy loss rates for both mechanisms are
\begin{equation}
    \dot{T}_{\mathrm{(sync, IC)},e} = \frac{4}{3} \beta^2 \gamma^2 c \sigma_{\rm T} U_{(B,\gamma)},
\end{equation}
where $\sigma_{\rm T}$ is the Thomson cross section, $U_B$ is the magnetic energy density, $U_\gamma$ is the radiation energy density, $\gamma$ is the electron Lorentz factor, and $\beta$ is the electron speed divided by $c$. By comparison, we can write the energy loss rate due to ionisations and bremsstrahlung -- the two processes whose rates are proportional to $\nH$ -- as
\begin{equation}
    \dot{T}_{\mathrm{(ion, brem)},e} = \nH \beta c \mathcal{L}_{\mathrm{(ion,brem)},e},
\end{equation}
where $\mathcal{L}_{\mathrm{ion},e}$ and $\mathcal{L}_{\mathrm{ion},e}$ are the loss functions for ionisation and bremsstrahlung, respectively. The former is given by an expression analogous to \autoref{eq:ion_loss_p}, but using the differential cross section for ionisations by electrons instead of protons, and with a maximum kinetic energy $W_{\rm max} = (T_e - I)/2$. The analogous expression for the bremsstrahlung loss function is
\begin{equation}
    \Lebrem =  \int \Eg \frac{d\sigma_{\mathrm{brem},e}}{d\Eg} \, d\Eg,
\end{equation}
where $d\sigma_{\mathrm{brem},e}/d\Eg$ is the differential cross section for production of photons of energy $\Eg$ by bremsstrahlung. Much of the energy loss occurs via photons whose energy is comparable to that of the CR, but for the purposes of the CSDA approximation, we adopt the expression $\Lebrem \approx (1/3) r_0^2 T_e$, where $r_0$ is the classical electron radius, which accurate to better than 40\% at electron energies $>1$ keV, and to better than 10\% at energies $>1$ MeV.

Given these expressions, we parameterise the importance of synchrotron and IC losses in terms of
\begin{equation}
    f_\mathrm{(sync,IC)} \equiv \frac{4\sigma_{\rm T} U_{(B,\gamma)}}{3\nH \Leionone},
    \label{eq:fsync}
\end{equation}
where $\Leionone = 1.04\times 10^{-17}$ eV cm$^2$ is the ionisation loss function evaluated at $p/m_e c=1$, where $p$ is the CR electron momentum;\footnote{At this energy, the loss functions for H~\textsc{i} and H$_2$ differ by $<1\%$, so we do not bother to distinguish them.} this quantity is, to order unity, the ratio of the (synchrotron, IC) loss rate to the ionisation loss rate at $p=m_e c$. With these definitions, we can express the total electron loss rate summed over all loss processes as $\dot{T}_e = \nH \beta c \mathcal{L}_e$, where
\begin{equation}
    \Le \equiv \Leion + \Lebrem + \beta \gamma^2 \left(f_{\rm sync} + f_{\rm IC}\right) \Leionone.
\end{equation}
Physically-realistic values of $f_{\rm sync}$ and $f_{\rm IC}$ in interstellar gas fall into a fairly narrow range -- both Milky Way-like conditions ($n_{\rm H} \approx 1$ cm$^{-3}$, $U_B \approx U_\gamma \approx 1$ eV cm$^{-3}$; \citealt{Draine11a}) and extreme starburst-like conditions ($n_{\rm H}\sim 10^3$ cm$^{-3}$, $U_B\sim U_\gamma \sim \mbox{few}$ keV cm$^{-3}$; \citealt{Krumholz20a}) give $f_\mathrm{(sync,IC)} \sim 10^{-7}$, simply because gas density, magnetic field, and interstellar radiation field intensity all tend to vary together. We will therefore adopt this as a fiducial value in what follows. This means that, as expected, synchrotron and IC losses are unimportant for low-energy CR electrons. However, since loss rates from both processes scale with energy as $\gamma^2$, while the ionisation loss function scales roughly as $\gamma^{-1}$, synchrotron and IC become increasingly important at higher energies. 

We can now proceed to calculate the ionisation and photon production rates as we did for protons. For ionisation, the total number of ionisations $N_{\mathrm{ion},e}$ produced by a CR electron with initial kinetic energy $T_{\mathrm{i},e}$ is given by \autoref{eq:Nion}, simply replacing the initial proton kinetic energy, ionisation cross section, and loss function with their equivalents for an electron, $T_{\mathrm{i},e}$, $\sie$, and $\Le$; the ionisation efficiency $\Psiie$ is defined analogously. Photon production at $\gamma$-ray energies and the photon production efficiency due to IC and bremsstrahlung, and $dN_{\gamma,e}/d\Eg$ is similarly given by \autoref{eq:dNdEgamma} and \autoref{eq:Psiep} with proton quantities replaced by electron ones, and the inelastic collision photon production differential cross section $d\sigma_{\gamma,p}/d\Eg$ replaced by the sum of the differential IC and bremsstrahlung cross sections, $d\sigma_{\mathrm{IC},\gamma,e}/d\Eg + d\sigma_{\mathrm{brem},\gamma,e}/d\Eg$. As with protons, we defer numerical evaluation to \autoref{ssec:criptic_numerical}.

\subsection{\textsc{criptic} simulations}
\label{ssec:criptic_numerical}

\subsubsection{Numerical method}

In order to calculate $N_{\mathrm{ion},p}$ and $E_{\gamma,p}$ numerically, without the approximations required by the CSDA, we carry out a series of simulations using the \textsc{criptic} CR propagation code. The full numerical setup for our simulations is provided in a public repository -- see the Data Availability statement for details. Each of our simulations consists of a monochromatic source of CR particles placed in a uniform medium of either molecular H$_2$ or atomic H~\textsc{i} with number density of H nuclei $\nH=10^3$ cm$^{-3}$, and magnetic and radiation fields chosen to have reasonable values for a starburst galaxy. Specifically, we set $f_{\rm sync} = f_{\rm IC} = 10^{-7}$; the corresponding magnetic and radiation energy densities are $370$ eV cm$^{-3}$, roughly the level expected for the midplane of a moderate starburst galaxy \citep[e.g.,][]{Krumholz20a, Crocker21a, Crocker21b}; the radiation field consists of the cosmic microwave background plus a dilute black body with a temperature of 40 K. We explore the effects of varying $f_{\rm sync}$ and $f_{\rm IC}$, and of varying the radiation temperature, in \aref{app:fsyncIC}. Since we are interested in the maximum number of ionisations and maximum $\gamma$-ray emission possible, we disable all CR transport by setting the CR diffusion coefficients and streaming speed to zero, so that no CRs escape. We carry out a total of 200 such simulations -- 50 each for sources injecting protons and electrons into fully atomic or fully molecular media. For the simulations where the source injects protons, the injected CR kinetic energies varying uniformly in logarithm between the pion production threshold $T_\pi = 0.28$ GeV and $10^6$ GeV; we choose the lower limit on our exploration to be $T_\pi$ because, below this limit, the only loss process for protons is ionisations, and the CSDA approximation is extremely accurate for this mechanism. For electrons, our energies are uniformly spaced from 100 MeV to $10^6$ GeV; again, the CSDA is extremely accurate for lower energies, since the loss processes that cannot be treated as continuous (and that \textsc{criptic} correctly treats as catastrophic) -- bremsstrahlung and IC scattering outside of the Thomson limit -- are unimportant compared to ionisation at energies below 100 MeV.

In the \textsc{criptic} simulations, we use a packet injection rate of $2\times 10^{-7}$ s$^{-1}$, a secondary production factor $f_{\rm sec} = 0.2$, and a step size control parameter $c_{\rm step} = 0.05$ -- see \citet{Krumholz22a} for precise definitions of these parameters. We follow CRs until their energies drop below 1 keV; below this energy, loss processes that are not included in \textsc{criptic} such as charge exchange cannot be neglected. However, as we will see, this choice has minimal effects, since CRs below this energy contribute negligibly to the total ionisation and $\gamma$-ray production budgets. We run each simulation for $10^{14}$ s; for comparison, the time required for the CR population to reach steady-state is of order the loss time $t_{\rm loss} = 1/n\sigma_{\rm inel} c \approx 10^{12}$ s, so the simulation time is long enough for the system to reach statistical steady state. We record the instantaneous specific $\gamma$-ray luminosity $dL_\gamma/dE_\gamma$ and ionisation rate $\dot{N}_{\rm ion}$ of the system at intervals of $5\times 10^{12}$ s (roughly 5 loss times) from $1.5\times 10^{13}$ to $10^{14}$ s, taking the mean of these samples as our estimate; the variance of the samples is in most cases $\sim 10-20\%$.  Dividing our estimates of the specific luminosity and ionisation rate by the CR injection rate then yield numerical estimates of $N_{\mathrm{ion},p}$ and $E_{\gamma,p}$, the number of ionisations and total energy radiated per injected CR proton, and similarly for electrons.

\subsubsection{Simulation results and comparison to the CSDA}

\begin{figure}
    \includegraphics[width=\columnwidth]{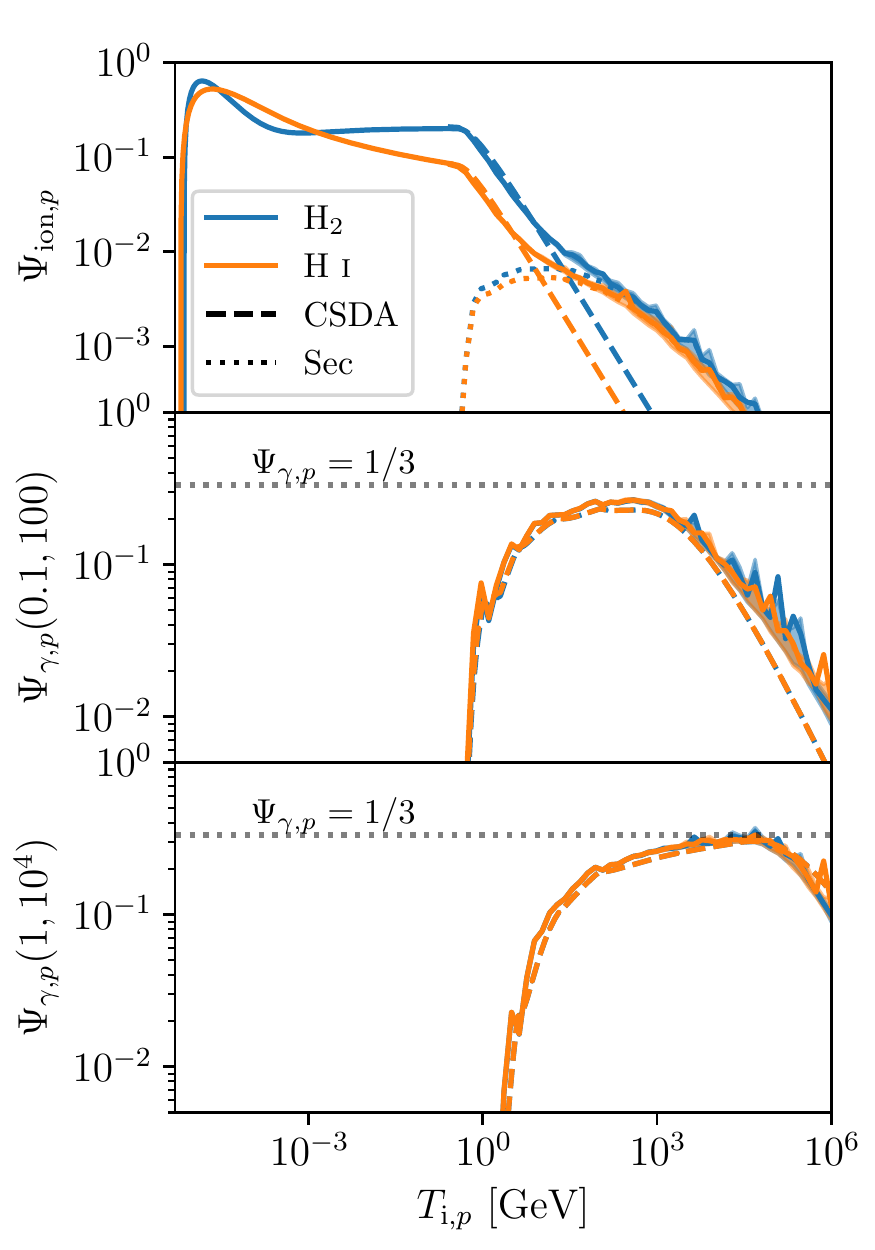}
    \caption{
    \label{fig:NionLp}
    Ionisation efficiency $\Psiip$ (top panel) and $\gamma$-ray production efficiencies $\Psiep$ for the band passes $(0.1,100)$ GeV (middle panel) and $(1,10^4)$ GeV (bottom panel) as a function of initial proton energy $\Tip$. Solid lines show the full numerical result, computed by averaging over time as described in the main text; the shaded band indicates the 16th to 84th percentile range of the variations. Dashed lines show approximate results obtained with the CSDA, and dotted lines in the top panel show ionisations due to secondaries, and computed from the mean of the numerical results. Blue lines show results for a molecular environment where all H is in the form of H$_2$, orange lines show results for an atomic environment where all H is in the form of H~\textsc{i}. The dotted lines in the lower two panels show $\Psiep = 1/3$, the upper limit corresponding to a proton that loses all its energy to pion production, and where the resulting neutral pions ultimately decay into $\gamma$-rays whose energies fall within the sensitivity range.
    }
\end{figure}

We plot $\Psiip$ and $\Psiep$ as functions of the initial proton energy in \autoref{fig:NionLp}; for the latter quantity we show the efficiencies computed over the interval $(E_0,E_1) = (0.1, 100)$ GeV (middle panel; roughly the energy range observed by \textit{Fermi}) and $(1,10^4)$ GeV (bottom panel; roughly the energy range to which CTA is sensitive for comparatively faint sources such as star-forming galaxies). We show both the full numerical results obtained using \textsc{criptic} and the CSDA approximation; for the latter, we use the cross sections computed exactly as in the full numerical results. We refer readers \citet{Krumholz22a} for full details, but to summarise here: we use the semi-analytic model of \citet{Rudd92a} to compute the total and differential proton ionisation cross sections, while our nuclear inelastic scattering cross section and corresponding differential photon production cross section come from \citet{Kafexhiu14a}, who provide analytic fits to the results of a large suite of particle Monte Carlo simulation results.

The plot shows that, for $\Tip$ from $\approx 0.1$ MeV to $\approx 1$ GeV, in molecular gas the efficiency $\Psiip \approx 0.2$ independent of energy, while in atomic gas it varies only weakly, going from $\approx 0.1 - 0.6$ over this energy range. The bump and then fall to zero at low energy occurs as we approach the kinematic threshold $(m_p/4 m_e)I$, while the downturn at higher energies occurs because, for protons above the pion production threshold $T_\pi = 0.28$ GeV, most energy goes into nuclear inelastic losses instead. In this regime, we approach $\Psiip\propto 1/\Tip$, with that scaling becoming almost exact in the CSDA, but a slightly flatter scaling once we account for the effects of ionisation by secondaries, which become dominant for $\Tip\gtrsim 10$ GeV. 

For $\gamma$-ray production, the results of the CSDA are very similar to those of the full numerical treatment at all energies, and the results are nearly identical for atomic or molecular background gas. Our results show that very close to $1/3$ of the losses through the nuclear inelastic channel are eventually radiated in the form of photons, as expected, since close to $1/3$ of the pions will be $\pi^0$ that subsequently decay into $\gamma$-rays. This leads to $\Psiep \approx 0.2 - 0.3$ over a broad range in energy for $\Tip \gtrsim 1$ GeV, the point at which nuclear inelastic losses begin to dominate. For a band pass of $0.1-100$ GeV, roughly corresponding to the sensitivity range of \textit{Fermi}, this relationship begins to break down at $\Tip\gtrsim 1$ TeV, as the photon emission shifts out of the energy band over which we are integrating. Similarly, for the $1-10^4$ GeV band pass corresponding roughly to CTA sensitivity, the relationship breaks down for protons with initial energies $\lesssim 10$ GeV due to photon emission at energies below the minimum energy to which the detector is sensitive. 

\begin{figure}
    \includegraphics[width=\columnwidth]{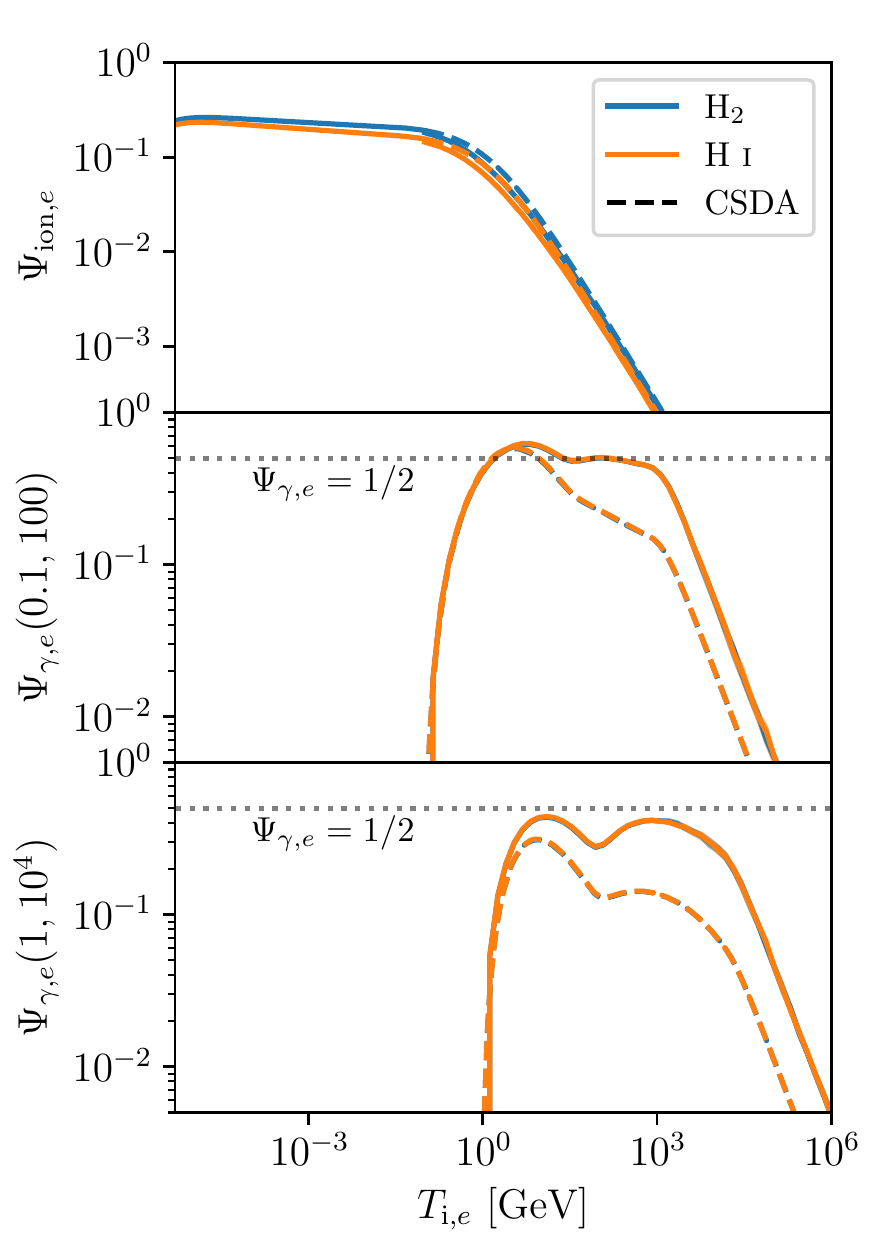}
    \caption{
    \label{fig:NionLe}
    Same as \autoref{fig:NionLp}, but for CR electrons rather than protons.
    }
\end{figure}

We show ionisation and $\gamma$-ray production efficiencies for electrons on \autoref{fig:NionLe}. For the CSDA, we again use the same microphysical cross sections as in the \textsc{criptic} simulations; in particular, the total and differential ionisation cross sections come from relativistic BEQ model of \citet{Kim00b}, while our expressions for bremsstrahlung, synchrotron, and IC emission follow the treatment of \citet{Blumenthal70a}. The numerical treatments of bremsstrahlung and IC scattering properly account for cases where the emitted photon energy is a large fraction of the electron energy (which, for IC, requires use of the full Klein-Nishina cross section rather than the Thomson approximation), and thus the CSDA is not applicable.

We see that the ionisation budget for electrons behaves qualitatively similarly to that of protons, in that for electron energies $\lesssim 1$ GeV most losses are into ionisation and $\Psiie$ is nearly constant.\footnote{The slight downturn in $N_{\mathrm{ion},e}$ at the lowest energies is an artefact of the minimum 1 keV at which we stop following CRs; however, as noted above, this has negligible effects on our calculation of the overall budget.} The results for atomic or molecular media differ only marginally. For larger initial energies, we find the same $\Psiie \propto 1/\Tie$ scaling as for protons, as other loss mechanisms dominate. Unlike for protons, the CSDA approximation remains nearly perfect in this regime, because secondaries are unimportant.

For $\gamma$-ray production, the pattern is slightly different. We again see that for electrons with initial energies that fall within the energy band pass of the detector ($0.1-100$ GeV or $1-10^4$ GeV), we have $\Psiie\approx 0.5$, i.e., half the energy is radiated as $\gamma$-rays within the observable range; the factor of two is because roughly half the energy is lost to synchrotron radiation, which emerges at lower energies. We see that the CSDA is reasonably accurate at energies up to $\approx 10$ GeV, but begins to under-predict the luminosity at higher energies, eventually reaching a factor of $\approx 5$ error at the highest energies, where inverse Compton scattering moves out of the Thomson regime and Klein-Nishina effects become important. As expected, results for atomic or molecular background media are nearly identical, since this distinction is only significant for ionisation losses, which are unimportant for CRs at the energies that produce $\gamma$-rays. \add{While cross sections per free particle obviously depend on the number of free particles per unit mass, the total fraction of the initial energy deposited in the various possible loss channels by a high-energy CR does not.}

\subsection{Spectral-averaged ionisation and $\gamma$-ray production efficiencies}
\label{ssec:spectral_average}

Our next step is to use the ionisation and $\gamma$-ray production budgets we have computed for individual CRs and convolve them with a spectrum of CRs injected with differing momenta. Let us suppose that CR protons are injected with a powerlaw spectrum of momentum over some momentum range $p_0$ to $p_1$, as suggested by both models of CR acceleration and observations of individual CR sources \citep[e.g.,][]{Caprioli12a, Bell13a}. The number of CR protons injected per unit time per unit momentum is therefore
\begin{equation}
    \frac{d\dot{n}_p}{dp} = \frac{\mathcal{N}}{m_p c} 
    \left\{
    \begin{array}{ll}
    x^{-q}, & x\in(x_0,x_1) \\
    0, & \mathrm{otherwise}
    \end{array}
    \right.,
    \label{eq:dndt}
\end{equation}
where for convenience we have defined $x = p/m_p c$ as the dimensionless proton momentum. It is convenient to express the normalisation in terms of the total (kinetic) luminosity of the CR proton injection, $L_{\add{\mathrm{CR},}p}$, in which case we have
\begin{equation}
    \int_{x_0}^{x_1} (\gamma-1) m_p c^2 \mathcal{N} x^{-q} \, dx = L_{\add{\mathrm{CR},}p},
\end{equation}
where $\gamma = \sqrt{1+x^2}$ is the CR proton Lorentz factor. Evaluating the integral gives
\begin{equation}
    \mathcal{N} = \frac{L_{\mathrm{CR},p}}{m_p c^2} \phi_p
\end{equation}
where $\phi_p$ is a dimensionless normalisation factor given by
\begin{equation}
    \phi_p = \left[
    \frac{x_0^{1-q}-x_1^{1-q}}{1-q}
    +
    \mathrm{B}_{c_0}(a,b) - \mathrm{B}_{c_1}(a,b)
    \right]^{-1},
\end{equation}
$a = q/2-1$, $b=(1-q)/2$, $c_{0,1} = (1 + x_{0,1})^{-2}$, and $\mathrm{B}_x(a,b)$ is the incomplete Beta function, $\mathrm{B}_x(a,b) = \int_0^x t^{a-1}(1-t)^{b-1}\,dt$.

Given the CR proton injection rate per unit momentum, we can compute the corresponding total rate at which CRs produce ionisations simply by integrating over the momentum distribution, and similarly for the $\gamma$-ray luminosity. Specifically, we have
\begin{eqnarray}
    \dot{N}_{\mathrm{ion},p} & = & \mathcal{N}  \frac{m_p c^2}{I} \int_{x_0}^{x_1} \Psiip (\gamma-1) x^{-q} \, dx \\
    L_{\gamma,p}\!(E_0,E_1) & = & \mathcal{N} m_p c^2 \int_{x_0}^{x_1} \Psiep (\gamma-1) x^{-q} \, dx,
\end{eqnarray}
where $L_{\gamma,p}\!(E_0,E_1)$ is the $\gamma$-ray luminosity emitted in the energy range from $E_0$ to $E_1$, and $\Psiip$ and $\Psiep$ are evaluated at initial kinetic energy $\Tip = (\gamma-1)m_p c^2$. We can, in turn, use these results to define spectrally-averaged ionisation and $\gamma$-ray production efficiencies
\begin{eqnarray}
    \label{eq:Psiipavg}
    \Psiipavg \equiv \frac{\dot{N}_{\mathrm{ion},p} I}{L_p} & = & \phi_p \int_{x_0}^{x_1} \Psiip (\gamma-1) x^{-q} \, dx \\
    \label{eq:Psiepavg}
    \Psiepavg \equiv \frac{L_{\gamma,p}}{L_p} & = & \phi_p \int_{x_0}^{x_1} \Psiep (\gamma-1) x^{-q} \, dx,
\end{eqnarray}
where we have omitted the explicit dependence of $\Psiepavg$ on $E_0$, $E_q$, and $q$ for compactness. We can of course define analogous expressions for CR electrons, simply replacing $x = p / m_p c$ with $y = p / m_e c$.

\begin{figure}
    \includegraphics[width=\columnwidth]{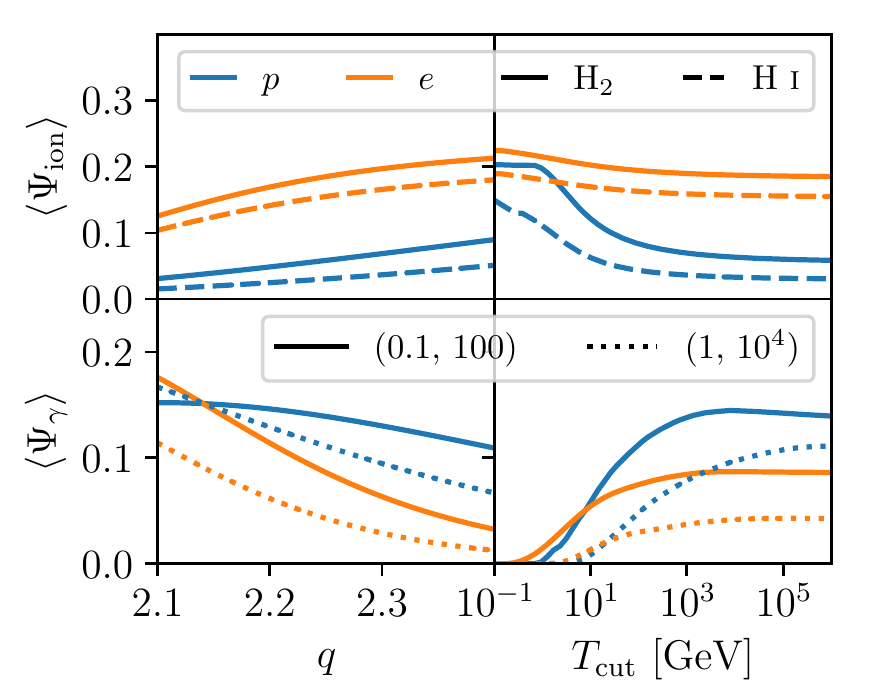}
    \caption{
    \label{fig:Psi}
    Spectral-averaged ionisation efficiency $\langle\Psi_\mathrm{ion}\rangle$ (top row) and $\gamma$-ray production efficiency $\langle\Psi_\gamma\rangle$ (bottom row) as a function of CR injection spectral index $q$ (left column) and cutoff energy $T_{\rm cut}$ (right column). Blue lines show protons, orange lines electrons. In the top row, solid lines show results for a pure H$_2$ background medium, dashed lines for a pure H~\textsc{i} medium. In lower row, solid lines correspond to $\gamma$-ray emission over a $(0.1, 100)$ GeV band pass, and dotted lines to a $(1, 10^4)$ GeV band pass.
    }
\end{figure}

In the left column of \autoref{fig:Psi} we plot \add{$\Psiipavg$ and $\Psiepavg$, and their electron equivalents,} as a function of $q$, using \add{lower and upper} limits on the injection distribution of $1$ keV and 1 PeV, respectively\footnote{\add{Note that the actual lower energy cutoff of the injection distribution is unknown, and our choice of 1 keV is arbitrary. However, this choice does not matter because the results are completely insensitive to the choice of lower energy cutoff as long as the cutoff is at highly sub-relativistic energies. This is because for a spectrum that is a powerlaw in momentum with a realistic spectral index, sub-relativistic CRs carry a negligible portion of the total CR luminosity budget. See \aref{app:differential_contributions} for details.}}; \add{we show results over the range $q=2.1$ to $2.4$, the plausible range for the ISM injection spectral index of CRs based on both observations and CR acceleration theory \citep[e.g.,][]{Caprioli12a, Bell13a}. I}n the right column we plot them as functions of $T_{\rm cut}$ for an injection energy range from 1 keV to $T_{\rm cut}$, for a spectral index $q=2.25$. \add{We also provide a more detailed investigation of which ranges of CR proton and electron energy make the largest contributions to these averages in \aref{app:differential_contributions}.} We find that the ionisation efficiency is largely insensitive to $q$ for both protons and electrons, with changes in index from $2.1$ to $2.4$ yielding only tens of percent differences. Ionisation efficiency is also insensitive to cutoff energy for electrons, since most of the available electron energy budget lies at energies where ionisation is dominant. For protons \add{in an H$_2$ background}, ionisation efficiency gradually decreases from $\approx \add{0.2}$ to $\add{\approx 0.06}$ as the cutoff energy increases from $T_{\rm cut} \sim 1$ to $\sim 10$ GeV and pion losses become significant \add{(solid blue line in the top right panel of \autoref{fig:Psi}); the efficiency is slightly lower in an H~\textsc{i} background, but the qualitative trend with $T_\mathrm{cut}$ is the same (dashed blue line in the top right panel of \autoref{fig:Psi})}. By contrast, $\gamma$-ray production efficiency is mostly insensitive to $q$ for protons, but somewhat sensitive for electrons, and for both protons and electrons it is insensitive to $T_{\rm cut}$ until $T_{\rm cut}$ comes within a factor of a few of the upper energy limit of the band pass. The figure also shows that CR electrons are $\approx 3\times$ more efficient than protons at producing ionisation and $\approx 2-3\times$ less efficient at producing $\gamma$-ray emission, depending on the band pass. These two results together mean that CR electrons will be subdominant for both ionisation and $\gamma$-ray production, since the total electron energy budget is expected. to be $\approx 10-20\%$ the proton energy budget \cite[e.g.,][]{Lacki10a}.

We \add{provide tabulated} values of $\Psiipavg$, $\Psiieavg$, $\Psiepavg$, and $\Psieeavg$ for some sample sets of parameters in \aref{app:tabulation}. In what follows, for convenience whenever we require numerical values we will use \add{efficiencies computed for the case}
$q=2.25$, $T_\mathrm{cut} = 10^6$ GeV\add{: $\Psiipavg = 0.058$, $\Psiieavg = 0.185$, $\Psiepavg = (0.139, 0.111)$, and $\Psieeavg = (0.086, 0.43)$, where the first number in parentheses is for the $(0.1,100)$ GeV band pass and the second for $(1,10^4)$ GeV}.

\section{Ionisation and diffuse $\gamma$-ray budgets of star-forming galaxies}
\label{sec:budgets}

Our next step is to determine the budgets for ionisation and diffuse $\gamma$-ray production in star-forming galaxies from the efficiencies we have computed. For this purpose we will consider a star-forming galaxy with total star formation rate $\dot{M}_*$ and gas mass $\Mg$, such that the gas depletion time $\tdep = \Mg/\dot{M}_*$. We consider a range of possible CR sources associated with star formation below. Generically, for any CR acceleration mechanism that is ultimately powered by star formation, we can express the energy budget for that mechanism in terms of $\langle E/M_*\rangle_\mathrm{m}$, defined as the total energy provided by that mechanism per unit mass of stars formed, averaging over the stellar initial mass function; thus for example $\langle E/M_*\rangle_\mathrm{SN}$ represents the total energy in supernova explosions per unit mass of stars formed. We similarly assign each mechanism proton and electron acceleration efficiencies $\epsilon_{\mathrm{m},p}$ and $\epsilon_{\mathrm{m},e}$, defined as the fraction of the energy provided by that mechanism that is ultimately deposited in non-thermal protons and electrons. Thus the total CR proton luminosity for any mechanism $m$ takes the generic form
\begin{equation}
    L_{p,\mathrm{m}} = \epsilon_{\add{\mathrm{m},p}} \dot{M}_* \left\langle \frac{E}{M_*}\right\rangle_\mathrm{m},
\end{equation}
and similarly for electrons.

From the CR luminosities, together with the efficiencies computed in \autoref{ssec:spectral_average}, we can compute the maximum number of primary ionisations per unit time each mechanism is capable of producing as 
\begin{equation}
    \dot{N}_{\mathrm{ion,m}} = \frac{\dot{M}_*}{I} \left\langle\frac{E}{M_*}\right\rangle_\mathrm{m} \epsilon_{\mathrm{m},p} \Psiipavg_\mathrm{m}
    \left(1 + \delta_{\mathrm{m}} \frac{\Psiieavg_\mathrm{m}}{\Psiipavg_\mathrm{m}}\right),
    \label{eq:Nion_gal}
\end{equation}
where $\delta_\mathrm{m} \equiv \epsilon_{\mathrm{m},e}/\epsilon_{\mathrm{m},p}$ is the ratio of electron to proton luminosity for that mechanism, and $\Psiipavg_\mathrm{m}$ and $\Psiipavg_\mathrm{m}$ are the proton and electron ionisation efficiencies for that mechanism, which are functions of the injected CR spectrum. The total $\gamma$-ray production budget integrated over some bandpass is given by \add{a very similar} expression,
\begin{eqnarray}
    L_{\gamma,\mathrm{m}} & = & \dot{M}_* \left\langle\frac{E}{M_*}\right\rangle_\mathrm{m} \epsilon_{\mathrm{m},p} \Psiepavg_\mathrm{m}
    \left(1 + \delta_{\mathrm{m}} \frac{\Psieeavg_\mathrm{m}}{\Psiepavg_\mathrm{m}}\right)
    \nonumber \\
    & \equiv & 
    \left\langle\frac{L_\gamma}{\dot{M}_*}\right\rangle_\mathrm{m}\dot{M}_*,
    \label{eq:Lgamma_m}
\end{eqnarray}
where the quantity $\langle L_\gamma/\dot{M}_*\rangle_\mathrm{m}$ is the $\gamma$-ray budget per unit star formation from a given mechanism.

For the purposes of interfacing with astrochemical models and comparing with observations, it is most convenient to express the ionisation budget as the primary ionisation rate per H nucleon. The total number of H nucleons in the galaxy is $M_\mathrm{g}/\mu_\mathrm{H} m_\mathrm{H}$, where $m_\mathrm{H}$ is the hydrogen mass, and $\mu_\mathrm{H}$ is the mean mass per H nucleon in units of $m_\mathrm{H}$; for the standard cosmological mix of $\approx 75\%$ H and $\approx 25\%$ He by mass, $\mu_\mathrm{H} \approx 1.4$. Thus the maximum ionisation rate that the CRs accelerated by a given mechanism can sustain is
\begin{eqnarray}
    \zeta_\mathrm{m} & = & \frac{\mu_{\mathrm{H}} m_\mathrm{H}}{\tdep I} \left\langle\frac{E}{M_*}\right\rangle_\mathrm{m} \epsilon_{\mathrm{m},p} \Psiipavg_\mathrm{m}
    \left(1 + \delta_{\mathrm{m}} \frac{\Psiieavg_\mathrm{m}}{\Psiipavg_\mathrm{m}}\right)
    \nonumber \\
    & \equiv & \frac{\langle\zeta \tdep\rangle_\mathrm{m}}{\tdep},
    \label{eq:zeta_m}
\end{eqnarray}
where $\langle \zeta\tdep\rangle_\mathrm{m}$ is the ionisation budget per unit star formation rate per unit gas mass (where $\tdep$ is the inverse of the star formation rate per unit gas mass).

It is important to keep in mind some caveats regarding $\zeta_\mathrm{m}$, which will be important in the discussion that follows. First, recall that $\zeta_\mathrm{m}$ is a galactic average; ionisation rates can of course be higher in the vicinity of CR sources, and lower far from them. Second, $\zeta_\mathrm{m}$ includes the effects of neither escape of ionising CRs from galaxies, nor diffusive reacceleration of CRs in the ISM; the former will lower ionisation rates compared to this estimate, while the latter will raise them. We return to these issues in \autoref{sec:discussion}.

We now proceed to estimate the budgets associated with individual mechanisms. For convenience we collect the coefficients $\langle\zeta\tdep\rangle_\mathrm{m}$ and $\langle L_\gamma/\dot{M}_*\rangle_\mathrm{m}$ for each mechanism in \autoref{tab:coefficients}.

\begin{table}
\begin{center}
\begin{tabular}{lcccc}
\hline\hline
Mechanism
& \multicolumn{2}{c}{$\log\langle L_\gamma/\dot{M}_*\rangle$}
& \multicolumn{2}{c}{$\log\langle\zeta\tdep\rangle$} \\
& \multicolumn{2}{c}{[$\mbox{erg s}^{-1}/(\mathrm{M}_\odot\mbox{ yr}^{-1})$]}
& \multicolumn{2}{c}{[$\mbox{s}^{-1}\mbox{ Gyr}$]} \\
& $(0.1,100)$ GeV & $(1,10^4)$ GeV & H~\textsc{i} & H$_2$
\\ \hline
Supernovae & $39.48$ & $39.37$ & $-16.29$ & $-16.12$ \\
Stellar winds & $39.09$ & $38.98$ & $-16.69$ & $-16.51$ \\
Protostars & $38.48$ & $38.16$ & $-16.73$ & $-16.78$ \\
H~\textsc{ii} regions & $36.90$ & $36.79$ & $-18.87$ & $-18.70$ \\
\hline
Sum & $39.66$ & $39.54$ & $-16.05$ & $-15.91$ \\
\hline \hline
\end{tabular}
\end{center}
\caption{$\gamma$-ray production and ionisation budgets for various mechanisms, computed using fiducial parameter choices. For a galaxy with total star formation rate $\dot{M}_*$ and depletion time $\tdep = \Mg/\dot{M}_*$, where $\Mg$ is the total gas mass, we have $\gamma$-ray luminosity $L_\gamma = \langle L_\gamma/\dot{M}_*\rangle \dot{M}_*$ and ionisation rate per H nucleon $\zeta = \langle\zeta\tdep\rangle/\tdep$. Units are chosen such that the value for $\langle L_\gamma/\dot{M}_*\rangle$ gives the $\gamma$-ray luminosity for a galaxy with a star formation rate of 1 M$_\odot$ yr$^{-1}$ in units of erg s$^{-1}$, and the value of $\langle\zeta\tdep\rangle$ gives the ionisation rate per H nucleon for a galaxy with a depletion time of 1 Gyr in units of s$^{-1}$. For $\langle L_\gamma/\dot{M}_*\rangle$, the two columns give values for $\gamma$-ray luminosity integrated over bandpasses of $(0.1,100)$ and $(1,10^4)$ GeV, respectively. For $\langle\zeta\tdep\rangle$, the two columns give ionisation budgets for a pure H~\textsc{i} and a pure H$_2$ background ISM, respectively.
\label{tab:coefficients}
}
\end{table}

\subsection{Supernovae and massive stellar winds}
\label{ssec:supernovae}

Supernovae (SNe) have long been thought to dominate the acceleration of CRs. To compute the SN energy budget, $\langle E/M_*\rangle_\mathrm{SN}$, we use the \textsc{slug} stellar population synthesis code \citep{da-Silva12a, Krumholz15b}, assuming a Solar metallicity population, and using a \citet{Chabrier05a} initial mass function (IMF), MIST stellar evolution tracks \citep{Choi16a}, and the models of \citet{Sukhbold16a} to determine which stars end their lives as type II SNe. We assume an energy of $10^{51}$ erg per SN. Under these assumptions, we find $\langle E/M_*\rangle_\mathrm{SN} = 6.5\times 10^{48}$ erg M$_\odot^{-1}$. If we further adopt our fiducial values for all efficiencies and normalise to $\epsn = 0.1$ and $\delta_\mathrm{SN} = 0.1$, then plugging into \autoref{eq:Lgamma_m} and \autoref{eq:zeta_m} gives the coefficients shown in \autoref{tab:coefficients}.

In addition to SNe at the ends of their lives, while they are alive massive stars also produce fast, radiatively-driven winds that produce shocks and can therefore accelerate CRs. We again use \textsc{slug} to compute $\langle E/M_*\rangle_\mathrm{w}$, using the ``Dutch`` stellar wind model as described by \citet{Roy21a}. We find $\langle E/M_*\rangle_\mathrm{w} \approx 2.6\times 10^{48}$ erg M$_\odot^{-1}$, so the total energy budget is $\approx 40\%$ of that for SNe. The expected maximum energy of CRs accelerated in wind shocks is at least as high as that for SNe, if not higher \citep[e.g.,][]{HESS-Collaboration15a, Morlino21a, Albert21a}, and thus the ionisation and $\gamma$-ray production efficiencies should be essentially the same as for SNe. Similarly, though the acceleration efficiency $\epw$ and electron-to-proton ratio $\depw$ have not been explored as much as for SNe, the fact that a large number of star clusters have now been detected in $\gamma$-rays \citep[e.g.,][]{HESS-Collaboration15a, Saha20a, Sun20a, Albert21a} suggests that the efficiency cannot be too small. We therefore adopt $\epw = 0.1$ and $\depw = 0.1$ as fiducial values as well. Inserting these choices into \autoref{eq:Lgamma_m} and \autoref{eq:zeta_m} gives the coefficients for stellar winds shown in \autoref{tab:coefficients}.

\subsection{Protostellar accretion and outflow shocks}

Both the shocks that occur on the surfaces of accreting protostars and the shocks produced when outflows from those accreting stars impact on the surrounding ISM are potential sites of CR acceleration \citep[e.g.,][]{Padovani15a, Padovani20a}. Both of these phenomena are ultimately powered by the release of gravitational potential energy from the accreting material, and thus the energy budget is fundamentally related to the gravitational potential at the surfaces of accreting protostars. \citet{Krumholz11e} shows that, due to the fact that protostars are generally fully convective, and have cores stabilised to a nearly fixed temperature by deuterium burning, this potential is nearly independent of accretion history or stellar mass, at least for stars with masses up to a few M$_\odot$, which do not exhaust their primordial deuterium supply until after they finish accreting. Since such low mass stars constitute the great bulk of the stellar mass, we can estimate the energy budget based on them; the surface potential is $\add{\xi} \approx 6\times 10^{47}$ erg M$_\odot^{-1}$, and we therefore have $\langle E/M_*\rangle_\mathrm{acc} \approx \add{\xi}$ for accretion.

For protostellar outflows, we adopt the parameterisation introduced in \citet{Cunningham11a}, whereby outflows ultimately carry away a fraction $f_m$ of the final stellar mass, launched at a speed that is a fraction $f_v$ of the Keplerian speed at the stellar surface, $v_K = \sqrt{\add{\xi}/2}$. Thus the mean protostellar outflow energy released per unit stellar mass formed is $(\add{\xi}/2) f_m f_v^2$. Observations of outflow momentum imply that the combination $f_m f_v \sim 0.3$ \citep[e.g.,][]{Richer00a, Cunningham11a} and theoretical models suggest $f_v \approx 1 - 3$. Thus we can write the total energy budget for protostellar accretion and outflow shocks together as
\begin{equation}
    \left\langle\frac{E}{M_*}\right\rangle_\mathrm{ps} \approx \left(1 + \frac{f_w}{2}\right) \add{\xi},
\end{equation}
where $f_w = f_m f_v^2 \approx 0.3 - 1$. 

The CR acceleration parameters are significantly more uncertain for jets and accretion shocks than for SNe. \citet{Araudo21a} use observations of synchrotron emission from massive protostellar jets to estimate a proton acceleration efficiency $\epsilon_{\mathrm{ps},p} \approx 0.05$ and an electron to proton ratio $\delta_\mathrm{ps} \sim 0.1$, but with very large systematic uncertainties; it is also unclear whether the efficiencies will be similar for low-mass protostars, which though less-luminous individually, dominate the total available energy budget due to their vastly greater mass. Similarly, \citet{Padovani15a} estimate a maximum CR energy from jet shocks of $\sim 10$ GeV for protons and $<1$ GeV for electrons, while \citet{Araudo21a} find somewhat higher values of $\sim 0.1$ TeV for protons. Given the various uncertainties, we will adopt as fiducial values $\epsilon_{\mathrm{ps},p} = \delta_\mathrm{ps}=0.1$ (i.e., the same parameters as for SNe), and ionisation and $\gamma$-ray production efficiencies $\Psiipavg = 0.1$, $\Psiieavg = 0.2$, $\Psiepavg=(0.1, 0.05)$, and $\Psieeavg = (0.05, 0.01)$ as fiducial estimates, where as usual the first figure in parentheses is for the $(0.1,100)$ GeV $\gamma$-ray bandpass, and the second for $(1,10^4)$ GeV. We also adopt a fiducial value $f_w = 1$ for the wind energy. Inserting these choices into \autoref{eq:Lgamma_m} and \autoref{eq:zeta_m} gives the coefficients shown in \autoref{tab:coefficients}. The numerical results show that, for our fiducial assumptions, protostellar jets and accretion shocks are subdominant by a factor of $\sim 3$ compared to SNe for ionisations, and by an order of magnitude for $\gamma$-ray emission. However, this does not mean they cannot be dominant locally -- a point to which we return below.

\subsection{H~\textsc{ii} region shocks}

\citet{Padovani19a} suggest that H~\textsc{ii} region shocks can accelerate CRs. To estimate the energy budget associated with such shocks, we begin by considering an ionising source with photon luminosity $S$ embedded in a uniform background medium with number density of H nuclei $n_\mathrm{H}$ prior to the start of H~\textsc{ii} region expansion. \citet[][equation 7.35]{Krumholz17b} show that a time $t$ after the H~\textsc{ii} region begins expanding, the energy carried by the shell bounding it is
\begin{equation}
    E_{\mathrm{sh}} = 8.1 \times 10^{47}\;t_6^{6/7} S_{49}^{5/7} n_2^{-10/7} T_{i,4}^{10/7}\mbox{ erg},
\end{equation}
where $t_6 = t/10^6$ yr, $S_{49} = S / 10^{49}$ photons s$^{-1}$, $n_2 = n_\mathrm{H}/100$ cm$^{-3}$, and $T_{i,4}$ is the temperature of the ionised gas in units of $10^4$ K. To estimate the ionisation budget, we evaluate using $t_6 \approx 4$, roughly the lifetime of the stars' large ionising fluxes. The total ionising photon budget per unit mass of stars formed is $\langle S/M_*\rangle = 6.3\times 10^{46}$ photons M$_\odot^{-1}$ \citep{Krumholz17b}, so if individual H~\textsc{ii} regions have ionising luminosities $S$, then one such region is formed per $159 S_{49}$ M$_\odot$ of stars formed. Thus the total energy in H~\textsc{ii} region shells per unit mass of stars formed is
\begin{equation}
    \left\langle \frac{E}{M_*} \right\rangle_\mathrm{H~\textsc{ii}} = 1.7\times 10^{46} n_2^{-10/7} T_{i,4}^{10/7} S_{49}^{-2/7}\mbox{ erg M}_\odot^{-1}.
\end{equation}
The ionised gas temperature $T_i$ cannot be too different from $10^4$ K, so in order for H~\textsc{ii} regions to have an energy budget competitive with that of SNe ($\langle E/M_*\rangle_\mathrm{SN} \approx 7\times 10^{48}$), we would require either $\nH \lesssim 1$ cm$^{-3}$ or $S \lesssim 10^{45}$ s$^{-1}$. The former possibility is ruled out because regions with densities that low are predominantly neutral or warm ionised medium, with temperatures high enough that H~\textsc{ii} regions do not create strong shocks when expanding into them, while the latter possibility is ruled out because it is far below the ionising luminosity of even a single O star. We therefore conclude that the H~\textsc{ii} region shock energy budget must be significantly below that for SNe. We will adopt $n_2 = T_4 = S_{49} = 1$ as fiducial values for our numerical estimates, but these choices will make relatively little difference to the total budget simply because they only affect a subdominant component.

To complete our estimate, we require the CR acceleration parameters for H~\textsc{ii} regions, which are poorly studied. \citet{Padovani19a} predict that the maximum CR energies are $\gtrsim 100$ GeV, in which case the ionisation and $\gamma$-ray production efficiencies should be comparable to those for SNe, but there are no predictions in the literature for either the total energy acceleration efficiency or the ratio of electron and proton luminosities. In the absence of information, we assume that these are the same as for SNe, i.e., $\epsilon_{p,\mathrm{H~\textsc{ii}}} = \delta_{\mathrm{H~\textsc{ii}}} = 0.1$. Doing so gives the ionisation and $\gamma$-ray production budgets listed in \autoref{tab:coefficients}.

\subsection{Sum over all mechanisms}
\label{ssec:sum_mechanisms}

Summing over all the mechanisms we have identified, and using the fiducial values listed in \autoref{tab:coefficients}, we arrive at a final estimate for the total CR ionisation budget associated with star formation. This is
\begin{equation}
    \zeta_\mathrm{tot} = (0.89, 1.2)\times 10^{-16}
    \left(\frac{\tdep}{\mbox{Gyr}}\right)^{-1}\;\mbox{s}^{-1},
    \label{eq:zeta_tot}
\end{equation}
where the first number in parentheses is for an ISM dominated by H~\textsc{i} gas, and the second for an ISM dominated by H$_2$. Of this budget, roughly 60\% comes from SNe, 20-25\% from stellar winds, and 15-20\% from protostellar accretion shocks and jets. Repeating this exercise for $\gamma$-rays gives
\begin{equation}
    L_{\gamma,\mathrm{tot}} = (4.57, 3.47)\times 10^{39} \left(\frac{\dot{M}_*}{\mathrm{M}_\odot\;\mbox{yr}^{-1}}\right)\;\mbox{erg s}^{-1}
    \label{eq:Lgamma_tot}
\end{equation}
as the total $\gamma$-emission budget, with the first number applying to a $(0.1,100)$ GeV bandpass, and the second a $(1,10^4)$ GeV bandpass. Of this budget, SNe contribute roughly $2/3$, stellar winds a bit under $1/3$, and protostellar shocks and jets about $5\%$.

It is worth noting that our fiducial ratio of maximum $\gamma$-ray luminosity to star formation rate is a factor of $\approx 2$ lower than that given by \citet{Kornecki20a} at equal star formation rate. At first this might seem surprising, particularly because we include CR acceleration mechanisms that \citet{Kornecki20a}, who consider only SNe, do not. However, this is more than outweighed by a number of other factors. The single largest one is the assumed number of SNe per unit mass of stars formed: \citeauthor{Kornecki20a} assume 1 SN per 83 M$_\odot$ of stars formed, whereas our calculation with \texttt{slug} \citep{da-Silva12a, Krumholz15b} gives one SN per 155 M$_\odot$; the difference is partly because we use a \citet{Chabrier05a} IMF while \citeauthor{Kornecki20a} use a \citet{Chabrier03a} IMF, and partly because \citeauthor{Kornecki20a} assume that all stars with birth masses $>8$ M$_\odot$ produce SNe, while we determine which initial stellar masses yield SNe from the state of the art models of \citet{Sukhbold16a}, which predict failed SNe over part of this mass range.\footnote{Both our estimate of the number of SNe per unit mass of stars formed and that of \citet{Kornecki20a} are consistent with Milky Way observational constraints, which imply a core collapse supernova rate of $1.2-2.1$ per century \citep{Rozwadowska21a}. The Milky Way star formation rate is $\approx 1.5-2$ M$_\odot$ yr$^{-1}$ \citep{Chomiuk11a, Licquia15a}, so \citet{Kornecki20a}'s estimate corresponds to a Milky Way core-collapse SN rate of $1.8-2.4$ per century, while our revised estimate gives $1.0-1.3$ per century.} A second contributor is that \citeauthor{Kornecki20a} adopt $\Psiep = 0.25$, compared to our fiducial $\Psiep = 0.13$; this is partly because they neglect ionisation losses, which are sub-dominant but not entirely negligible at $\sim$GeV proton energies, and partly because they use older $\gamma$-ray production cross sections from \citet{Kelner06a}, which assume the ultra-relativistic limit, whereas we use the more recent result from \citet{Kafexhiu14a} that does not rely on the ultra-relativistic assumption; \citeauthor{Kelner06a} predict substantially more $\gamma$-ray production at $\lesssim 1$ GeV energies (e.g., see Figure 12 of \citeauthor{Kafexhiu14a}), leading to higher $\Psi_{\gamma,p}$ in the \textit{Fermi} band. A final contributing factor is that \citeauthor{Kornecki20a} assume that 10\% of SN energy goes into CR protons with energies $>1.2$ GeV, whereas our $\epsilon_p$ is the acceleration efficiency integrated over all proton energies; for our fiducial $q=2.25$ spectral index, \citeauthor{Kornecki20a}'s normalisation corresponds to $\epsilon_p = 0.133$.

\section{Discussion}
\label{sec:discussion}

We now examine some of the implications of our findings, both in the Milky Way and in other galaxies.

\subsection{Application to the Milky Way}

The average gas depletion time of the Milky Way is $\approx 3$ Gyr \citep{Licquia15a}, varying with galactocentric radius from $\approx 2$ Gyr in the H$_2$-dominated regions at $R \lesssim 5$ kpc, to $\approx 5$ Gyr near the Solar circle \citep{Kennicutt12a}. From \autoref{eq:zeta_tot}, this implies a mean primary ionisation budget $\zeta \approx 2-5\times 10^{-17}$ s$^{-1}$. This is a factor of at least a few lower than the mean value of $\approx 2\times 10^{-16}$ s$^{-1}$ inferred from astrochemical measurements in molecular clouds (\citealt{Indriolo12a, Indriolo15a, Porras14a, Zhao15a, Bacalla19a}; for recent reviews see \citealt{Padovani20a} and \citealt{Gabici22a}) and is more consistent with the value of $\approx 1-2\times 10^{-17}$ s$^{-1}$ implied by in situ measurements from \textit{Voyager} \citep{Cummings16a}.\footnote{The astrochemical measurements are likely to be revised down slightly, since \citet{Ivlev21a} have shown that the ratio of secondary to primary ionisations is $\approx 50\%$ larger than assumed in the past; since the astrochemical measurements are sensitive to the total ionisation rate, this will lead to a downward revision of the primary ionisation rate by $\approx 20\%$. However, this is a relatively minor difference.} Moreover, recall that \autoref{eq:zeta_tot} is the budget assuming all injected CRs give up all their energy inside the neutral medium of the galaxy; energy losses in ionised gas or via escape into the Galactic halo will reduce the ionisation budget. Indeed, the fact that the measured ionisation rate is close to the upper limit strongly suggests that the Milky Way is \textit{not} transparent to the low-energy CRs that dominate ionisation, as some authors have assumed \citep[e.g.,][]{Papadopoulos10a, Bisbas15a, Bisbas17a}. 

The situation for the $\gamma$-ray budget is far different: from \autoref{eq:Lgamma_tot} together with the Milky Way's inferred star formation rate of $\approx 1.5-2$ M$_\odot$ yr$^{-1}$ \citep{Chomiuk11a, Licquia15a}, the predicted $\gamma$-ray budget of the Milky Way in the (0.1,100) GeV band is $6-8\times 10^{39}$ erg s$^{-1}$, as compared to the observed value $\approx 8\times 10^{38}$ erg s$^{-1}$ \citep{Strong10a}. This discrepancy has long been known and can be accommodated naturally if the Milky Way is only $\sim 10\%$ calorimetric for CR protons \citep[e.g.,][]{Lacki11a, Kornecki20a, Crocker21a, Crocker21b}. Thus we are led to a picture in which the $\sim 0.1$ GeV protons responsible for most ionisations are largely calorimetric, while the $\sim 10$ GeV protons that dominate $\gamma$-ray production (c.f.~\autoref{fig:dPsidlnT}) are only $\sim 10\%$ calorimetric.

We can provide an independent cross-check on this picture by comparing the CR spectral shape observed locally to the shape expected for full calorimetry, which we compute using the CSDA for simplicity. Consider a kinetic energy interval from $T$ to $T+dT$; if the Galaxy is fully calorimetric, then every CR injected with initial energy $T_i > T$ will eventually pass through this interval, taking a time $dt = dT/\dot{T}$ to do so. Thus if CRs with initial energies $T_i > T$ are injected into the Galaxy at a rate $\dot{N}(>T)$, in steady state the total number of CRs in the Galaxy per unit energy $dT$ is
\begin{equation}
    \frac{dN}{dT} = \frac{\dot{N}(>T)}{\dot{T}} = \frac{\dot{N}(>T)}{n_\mathrm{H}\beta c \mathcal{L}},
\end{equation}
where $n_\mathrm{H}$ is the number density of H nuclei and $\mathcal{L}$ is the loss function. We can compute the injection rate $\dot{N}(>T)$ simply by integrating over the injection spectrum (\autoref{eq:dndt})
\begin{equation}
    \dot{N}(>T) = \int_{x_T}^{x_1} \mathcal{N} x^{-q} \, dx,
\end{equation}
where $x_T$ and $x_1$ are the dimensionless momenta corresponding to kinetic energy $T$ and to the maximum kinetic energy produced by the acceleration process, respectively. If the CRs are distributed over a volume $V$ in the Galaxy, and we assume that their directions are isotropic, then we can express the CR intensity per unit energy per unit solid angle as
\begin{equation}
    j = \frac{\beta c}{4\pi V}\frac{dN}{dT} = \frac{\dot{N}(>T)}{4\pi n_\mathrm{H} V \mathcal{L}}. 
\end{equation}
Note that $j$ depends on kinetic energy only via the injection spectrum and the loss function, so these two factors alone determine the spectral shape.

\begin{figure}
    \includegraphics[width=\columnwidth]{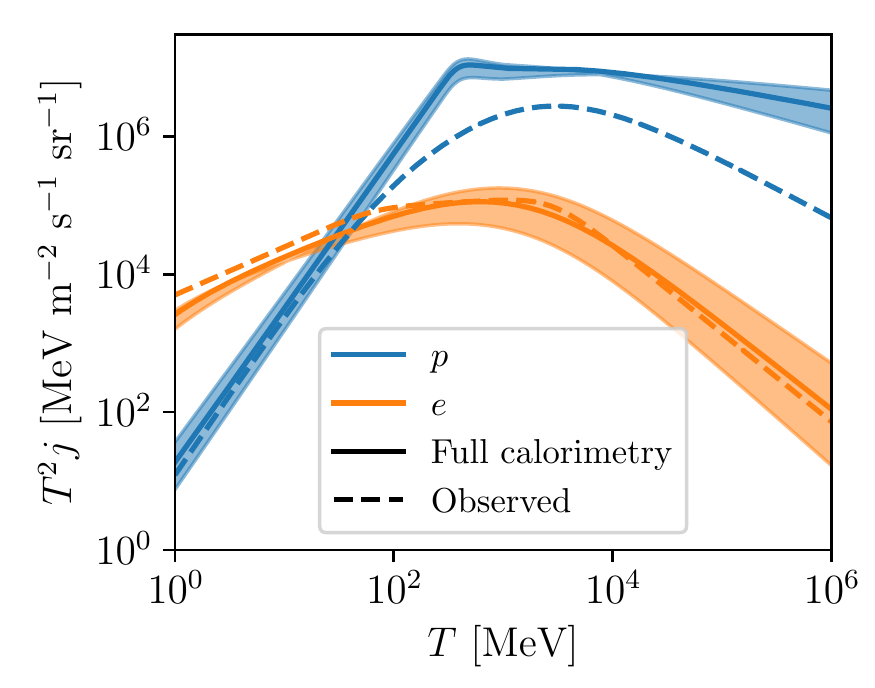}
    \caption{
    Comparison between the observed local interstellar spectrum $j$ of CR protons (blue) and electrons (orange) and spectra predicted under the assumption that the Galaxy is fully calorimetric. Dashed lines show the empirical fits to the observed LIS provided by \citet[his equations 14 and 15]{Gabici22a}, while solid lines and shaded bands show the LIS expected if the Galaxy is fully calorimetric, computed as described in the text. The central solid line is for an injection spectrum with index $q=2.25$, and the shaded band shows the results of varying $q$ over the range $2.1-2.4$.
    \label{fig:fluxspectra}
    }
\end{figure}

We plot $j$ as a function of $T$ for CR protons and electrons in \autoref{fig:fluxspectra}, using the loss function $\mathcal{L}$ for an H~\textsc{i} background; results for an H$_2$ background are very similar. For the purpose of setting the normalisation we adopt $n_\mathrm{H} = 1$ cm$^{-3}$ and a volume $V$ corresponding to a cylinder with a radius of 10 kpc and a half-height of 1 kpc, and we compute the injection rate including all the contributions listed in \autoref{sec:budgets} and using a fiducial spectral index $q=2.25$. For comparison we also plot the fits provided by \citet{Gabici22a} to the observed local interstellar spectra (LIS) of CR protons and electrons. The plot shows excellent agreement between the measured LIS and the optically thick predictions for electrons at all energies, and for protons at energies $\lesssim 0.1$ GeV. The agreement in normalisation is not particularly significant -- while our choices of $n_\mathrm{H}$ and $V$ are reasonable, clearly it would also be reasonable to adopt values that differ from our choices by factors of several. Instead, the important part of this plot is how the shapes of the predicted and observed spectra compare. For high proton energies we find that the optically thick assumption leads to a spectrum that is significantly shallower than the observed one, consistent with the conventional picture that a substantial fraction of high-energy CRs escape the Galaxy, and that the escape fraction increases with CR energy. By contrast, the agreement in spectral shape for low-energy protons, and for electrons of all energies, implies either that the Galaxy must be fully calorimetric for these CRs, or that any escape is energy-independent. Our cross-check against the shape of the LIS is therefore consistent with the quantitative conclusions we draw from our budget calculations.

Given this encouragement that our budgets are reliable, there does appear to be a real tension between the inferred ionisation budget and the ionisation rates inferred from astrochemical analysis of molecular clouds. We next consider three possible paths to resolving this tension.

\subsubsection{Non-uniform ionisation rates}

One possible solution is to consider that the astrochemical measurements may not be reflective of the true Galactic average. These measurements necessarily target molecular clouds, which may contain a significant number of local CR sources (driven by protostellar outflows, H~\textsc{ii} regions, or wind shocks, as considered in \autoref{sec:budgets}) that elevate their ionisation rate above the Galactic mean. As a simple thought experiment, if one were to hypothesise that CRs injected by SNe produce ionisation distributed uniformly over all neutral gas in the Galaxy, but those injected by stellar winds and protostars produce ionisations almost exclusively within molecular clouds, then the ionisation budget within molecular clouds would, for our fiducial parameters, increase to $\zeta_\mathrm{mc} = (0.75+0.47/f_\mathrm{mc})\times 10^{-16} (t_\mathrm{dep}/1\mbox{ Gyr})^{-1}$, where $f_\mathrm{mc}$ is the mass fraction in molecular clouds. Since $f_\mathrm{mc} \sim 0.1-0.5$ depending on the galactocentric radius over which one computes the average \citep{Kennicutt12a}, this implies an ionisation rate in molecular clouds of $1.7-5.5\times 10^{-16}$ s$^{-1}$, in good agreement with the astrochemically-inferred molecular cloud ionisation rates. If there are additional local sources in molecular clouds beyond those we have considered, for example magnetic reconnection events \citep{Gaches21a}, then there is additional room for the non-SN sources not to be so concentrated in molecular clouds or for some level of CR escape from the Galaxy. Conversely, however, this hypothesis depends crucially on the still poorly-understood details of CR transport around molecular clouds. Simulations suggest that the transport is complex and yields ionisation rates that are highly spatially variable \citep{Fitz-Axen21a}, and it is not clear if the ionisation budget supplied by sources within molecular clouds can be confined to the cloud volume. \add{Alternatively, significant spatial variations in the ionization rate could also be produced if the supernova sources are not distributed uniformly \citep{Phan2021,Phan2022}.}

\subsubsection{Type Ia supernovae}

We have focused on the contribution of CRs that trace star formation, but in the Milky Way type Ia SNe, which trace the older stellar population, occur at a rate comparable to core collapse SNe, and should accelerate CRs as efficiently as core collapse SNe. Quantitative estimates of the SNIa rate vary from $\approx 0.4$ per century \citep{Ruiter09a} to $\approx 1.4$ per century \citep{Adams13a}, compared to the $1.0-1.3$ core collapse SNe per century we estimate using \textsc{slug} together with the measured Galactic star formation rate (\autoref{ssec:sum_mechanisms}). Moreover, the mean energy release from SNIa is expected to be a factor of $\approx 1.5-2$ larger than for core collapse SNe \citep[e.g.,][]{Thielemann04a, Pakmor22a}. Thus SNIa likely provide a CR luminosity comparable to or even a factor of a few larger than the core collapse SNe that trace Galactic star formation.

What is less certain is how much ionisation or $\gamma$-ray emission these CRs will provide. A crucial difference between SNIa and core collapse SNe is that, because the former occur in an old stellar population, they tend to occur further from the Galactic plane. For external galaxies, \citet{Hakobyan17a} find that the scale height of core collapse SNe is comparable to that of the thin stellar disc, while the scale height of SNIa is a factor of $\approx 2-3$ larger. Thus while most core collapse SNe will at least initially deposit their CRs into relatively dense neutral gas near the Galactic plane, only a $\approx 1/2 - 1/3$ of SNIa will do so. Those SNe that occur well off the plane seem unlikely to produce much ionisation or $\gamma$-ray emission, since the CR protons they accelerate would need to diffuse or stream back toward the dense gas in the plane in order to do so. Even with this caveat, however, it is plausible, given the available energy budget, that SNIa in the Milky Way could produce a CR ionisation and $\gamma$-ray budget comparable to that of core collapse SNe. If so, this would go some distance to alleviating the ionisation rate tension. However, we emphasise that while this may be true of the Milky Way, it will not be for many other star-forming galaxies. The Milky Way is a green valley galaxy on the verge of quenching \citep{Bland-Hawthorn16b}, so its specific star formation rate is quite low, implying a ratio of type Ia to core collapse SNe higher than that expected for most star-forming galaxies.

\subsubsection{Second-order Fermi acceleration}
\label{sssec:fermi2}

A third possible solution would be to consider the contribution of second-order Fermi acceleration to the ionisation budget, as proposed by \citet{Drury17a}. Diffusion in momentum space with a diffusion coefficient $K_{pp}$ will cause particles with momentum $p$ to gain momentum at an average rate $\dot{p}_\mathrm{2F} = (2 + \alpha) K_{pp}/p$, where $\alpha \equiv d\ln K_{pp}/d\ln p$. The value and energy dependence of $K_{pp}$ are very poorly known, and are tied up in the question of whether CRs are self-confined, in which case the turbulence with which they interact is highly imbalanced, or externally confined, in which case it is likely close to balanced; the former scenario implies much less efficient acceleration than the latter \citep{Zweibel17a, Hopkins22a, Bustard22a}. \citeauthor{Drury17a} estimate that re-acceleration increases the CR luminosity of the Milky Way by $\approx 50\%$, but this result assumes external turbulence rather than self-confinement, which seems improbable for the $\lesssim\mbox{GeV}$ energies that dominate ionisation \citep[e.g.,][]{Xu16a, Zweibel17a, Krumholz20a, Kempski22a}. The \citeauthor{Drury17a} result also relies on a numerical value for the spatial diffusion coefficient that may be a significant overestimate if, as \citet{Sampson22a} suggest, the empirically-inferred diffusion coefficient in fact reflects transport by streaming coupled with turbulent motion of the underlying medium, rather than true microphysical diffusion. Conversely, however, \citeauthor{Drury17a}'s estimate is also obtained using the spectrum of CRs measured by \textit{Voyager}. If this is an underestimate of the Galactic average, that would imply a significantly larger energy contribution by second-order Fermi acceleration, since the rate of energy gain by this process is proportional to the CR number density. 

Given the uncertainties, it is difficult to make a convincing estimate of the contribution of second-order Fermi acceleration to the total ionisation budget. However, it is nonetheless an interesting exercise to ask whether second-order Fermi acceleration plausibly has the characteristics that would be required to explain the tension between the ionisation budget, the $\gamma$-ray budget, and the astrochemical measurements. To make this estimate we follow the approach of \citet{Recchia19a} by comparing the loss and gain timescales; for second-order Fermi acceleration to be able to add significantly to the ionisation budget, it must be able to increase particle energies on timescales similar to or faster than those on which they lose energy ($t_\mathrm{gain} \lesssim t_\mathrm{loss}$), since otherwise there will not be time for significant energy input to occur. We define the loss time as $t_\mathrm{loss} = T/\dot{T}_{\mathrm{loss}}$, where $\dot{T}_\mathrm{loss}$ is summed over all loss mechanisms. 

To compute the gain time, we note that the natural scaling expected between the diffusion coefficient in position space $K_{xx}$ and that in momentum space is $K_{pp} \approx \eta p^2 v^2 / K_{xx}$, where $\eta$ is a numerical factor $\sim 0.1$ for balanced turbulence but much smaller for unbalanced turbulence, and $v$ is the characteristic velocity of the turbulence responsible for acceleration -- either the Alfv\'en speed for diffusing CRs, or the flow speed for non-resonant acceleration of streaming CRs. Thus the gain time is
\begin{equation}
    t_\mathrm{gain} = \frac{T}{\dot{p}_{2F} (dT/dp)} \approx \left(\frac{d\ln p/d\ln T}{2+\alpha}\right)\frac{K_{xx}}{\eta v^2},
\end{equation}
and the condition for the gain time to be shorter than the loss time becomes
\begin{equation}
    K_{xx} \lesssim \left(\frac{2+\alpha}{d\ln p/d\ln T}\right) \eta v^2 t_\mathrm{loss}.
\end{equation}

\begin{figure}
    \includegraphics[width=\columnwidth]{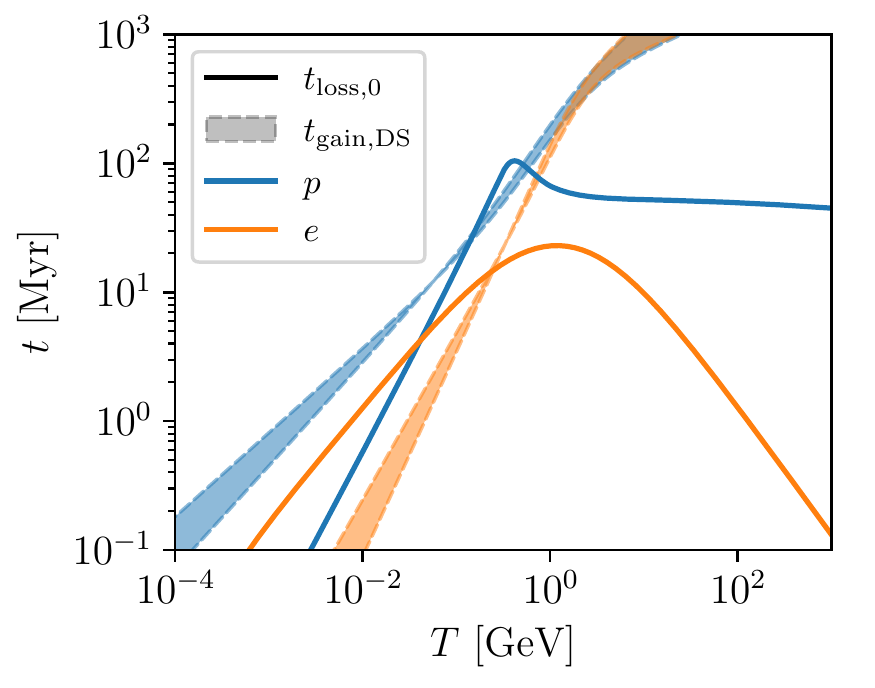}
    \caption{
    Loss and gain timescales as a function of particle energy. Solid lines show the loss timescales $t_\mathrm{loss,0}$ at the mean Milky Way ISM density $n_\mathrm{H}=1$ cm$^{-3}$, evaluated for protons (blue) and electrons (orange). Shaded regions correspond to the gain times $t_\mathrm{gain,DS}$ produced by the second-order Fermi acceleration model of \citet{Drury17a}, with the shaded region corresponding to the results of varying their parameter $\delta$, with $K_{xx}\propto p^\delta$, over their preferred range $\delta=0.3-0.6$.
    \label{fig:tloss}
    }
\end{figure}

We plot loss and gain times for protons and electrons as a function of kinetic energy in \autoref{fig:tloss}; for the loss times we scale to $n_\mathrm{H}=1$ cm$^{-3}$, roughly the mean density of the Milky Way's ISM. We therefore plot $t_\mathrm{loss,0}$ defined such that $t_\mathrm{loss} = t_\mathrm{loss,0} (n_\mathrm{H}/1\mbox{ cm}^{-3})$. The loss times shown are for H~\textsc{i}, since this is the dominant volume-filling medium in the Milky Way, but the results for H$_2$ are very similar. To give an example of gain times, we plot the \citet{Drury17a} model, which has $v = 30$ km s$^{-1}$ and $K_{xx} = 1.0\times 10^{28} \beta (p/m_p c)^\delta$ cm$^2$ s$^{-1}$ for protons, with $\eta = 4/[3\delta(4-\delta^2)(4-\delta)]$ and $\delta=0.3-0.6$; we compute gains for electrons by assuming that $K_{xx}$ is the same for protons and electrons of equal rigidity. 

Examining the figure, we can see two regimes where second-order Fermi acceleration could be significant. For protons, the loss time reaches a maximum value $t_\mathrm{loss,0} \approx 100$ Myr at $T_p \approx 0.4$ GeV, which is also in the kinetic energy range that contributes most strongly to ionisation (c.f.~\autoref{fig:dPsidlnT}). The loss time is only a factor of $\sim 2$ shorter at higher energies, but these CRs contribute little to ionisation, while the loss time is much shorter at lower energies ($t_\mathrm{loss}\sim T_p^{1.4-1.5}$), making these CRs hard to re-accelerate before their energy is drained by ionisation losses. Thus if re-acceleration of protons is to contribute significantly to the Galactic ionisation budget, it must be re-acceleration of $\sim 0.1-1$ GeV protons, since these are in the sweet spot where they can contribute to ionisation but do not suffer such rapid ionisation losses that they give up all their energy before there is an opportunity to re-accelerate them. The \citeauthor{Drury17a} model we plot does predict that second-order Fermi acceleration is significant in this energy range, since $t_\mathrm{gain} \lesssim t_\mathrm{loss}$.

For electrons, the loss time has a maximum of $\approx 20$ Myr at $T_e\approx 1$ GeV, but these high-energy electrons make relatively little contribution to the ionisation budget. The electron loss time is shorter for lower energy electrons, but $t_\mathrm{loss}$ varies with energy less steeply than for protons. Consequently, second-order Fermi acceleration for electrons is conceivably important at $\sim$MeV energies, which do contribute significantly to the ionisation budget. Indeed, the \citeauthor{Drury17a} model we plot naturally predicts significant re-acceleration for electrons in this energy range. However, we remind readers that for both protons and electrons this result is critically dependent on assuming balanced turbulence at the small length scales resonant with sub-GeV particles, contrary to theoretical expectation. If $\eta \ll 0.1$, as expected for unbalanced waves, then second-order Fermi acceleration is unlikely to be important.

Moreover, as pointed out by \citet{Recchia19a}, the energetic requirements associated with maintenance of a significant low-energy CR population (produced by second-order Fermi acceleration or any other mechanism) are formidable. Indeed, our calculation of the ionisation efficiency allows us to make this point even more strongly. The energy input per unit time required to sustain a mean primary CR ionisation rate per H nucleon $\zeta$ in gas with total mass $M_\mathrm{g}$ is $L_\mathrm{ion} = \zeta I \Mg / (\mu_\mathrm{H} m_\mathrm{H} \langle\Psi_{\mathrm{ion}}\rangle)$, and we can compare this to the total turbulent power provided by type II SNe, $L_\mathrm{SN,turb} = \epsilon_\mathrm{SN,turb} \dot{M}_* \langle E/M_*\rangle_\mathrm{SN}$, where $\epsilon_\mathrm{SN,turb}$ is the fraction of supernova energy that is ultimately injected into ISM turbulence (as opposed to being lost radiatively while supernova remnants are still expanding). The ratio is
\begin{eqnarray}
    \frac{L_\mathrm{ion}}{L_\mathrm{SN,turb}} & = &
    \frac{\zeta I \tdep}{\mu_\mathrm{H} m_\mathrm{H} \epsilon_\mathrm{SN,turb}\langle E/M_*\rangle_\mathrm{SN}\langle\Psi_{\mathrm{ion}}\rangle} 
    \label{eq:F2budget}
    \\
    & = & (0.36, 0.40) \zeta_{-16} \left(\frac{\epsilon_\mathrm{SN,turb}}{0.1}\right)^{-1} \left(\frac{\langle\Psi_\mathrm{ion}\rangle}{0.25}\right)^{-1} \left(\frac{\tdep}{\mbox{Gyr}}\right),
    \nonumber
\end{eqnarray}
where $\zeta_{-16} = \zeta/10^{-16}$ s$^{-1}$. As usual, the first number in parentheses is for H~\textsc{i} and the second for H$_2$. We have normalised $\langle\Psi_\mathrm{ion}\rangle$ to $0.25$, roughly the maximum efficiency we find at any energy (c.f.~\autoref{fig:NionLp} and \autoref{fig:NionLe}), and we have normalised the efficiency for conversion of SN energy to turbulence to 0.1, which is the maximum achieved by optimally-clustered SNe \citep{Gentry17a}; single SNe are a factor of $\approx 5$ less efficient, and realistic estimates of the mean efficiency are probably well below $0.1$.

The striking result is that, even with these generous scaling choices, \autoref{eq:F2budget} implies that achieving a mean ionisation rate of $\zeta\approx 1-2\times 10^{-16}$ s$^{-1}$ in a galaxy like the Milky Way with $\tdep$ of a few Gyr requires conversion of more than 100\% of the available turbulence produced by SNe into second-order Fermi acceleration. That is, even if one were to posit that the only mechanism by which interstellar turbulence in the Milky Way damps is by accelerating low-energy CRs, which then go on to ionise neutral gas as efficiently as possible, SN-driven turbulence would still not provide enough power to sustain mean ionisation rates as high as those found in Milky Way molecular clouds. While there are other power sources for interstellar turbulence -- radial transport of gas through the disc \citep{Krumholz18b} and cosmological accretion \citep{Ginzburg22a, Forbes22a} -- neither of those alternative sources are expected to be dominant in a low-redshift, gas-poor galaxy like the Milky Way. Consequently, our analysis echoes the conclusion of \citet{Recchia19a}: the hypothesis that an unseen population of low-energy CRs could sustain a mean Galactic ionisation rate as high as that inferred to exist in molecular clouds can be ruled out on energetic grounds.

\subsection{Budgets in external galaxies}

We can also use our models to estimate ionisation rates and calorimetry fractions in external galaxies. We make use of the star formation rate and $\gamma$-ray data compiled by \citet{Kornecki20a}, omitting the four galaxies from their sample -- NGC 2403, NGC 3424, NGC 4945, and Circinus -- that they conclude likely suffer from significant AGN contamination, combined with gas masses taken from a variety of sources to enable us to compute delpletion times. We list our sample galaxies in \autoref{tab:galaxies}. We then compute the calorimetry fraction of each galaxy as
\begin{equation}
    f_\mathrm{cal} = \frac{L_\gamma/\dot{M}_*}{\left\langle L_\gamma/\dot{M}_*\right\rangle},
    \label{eq:fcal}
\end{equation}
with the numerical value of the denominator given by \autoref{eq:Lgamma_tot}, and the primary ionisation rate budget of each galaxy, derived from its star formation rate, from \autoref{eq:zeta_tot} assuming the case of an H$_2$-dominated medium. We list this quantity in the Table as $\zeta_{\dot{M}_*}$.

\begin{table*}
    \begin{tabular}{lccccccc}
    \hline\hline
    Name & $\log \dot{M}_*$ & $\log\Mg$ & $\log L_\gamma$ & $\log \tdep$ & $\log f_\mathrm{cal}$ & $\log \zeta_{\dot{M}_*}$ & $\log\zeta_{L_\gamma/\Mg}$ \\
    & [M$_\odot$ yr$^{-1}$] & [M$_\odot$] & [erg s$^{-1}$] & [yr] & & [s$^{-1}$] & [s$^{-1}$]
    \\ \hline
    \multicolumn{8}{c}{Normal galaxies}
    \\ \hline
    SMC & $-1.57 \pm 0.05$ & $\phantom{0}8.51 \pm 0.30$ & $37.10 \pm 0.05$ & $10.08 \pm 0.31$ & $-0.99 \pm 0.07$ & $-16.99 \pm 0.31$ & $\cdots$ \\
    LMC & $-0.70 \pm 0.07$ & $\phantom{0}8.73 \pm 0.30$ & $37.77 \pm 0.06$ & $\phantom{0}9.43 \pm 0.31$ & $-1.19 \pm 0.09$ & $-16.34 \pm 0.31$ & $\cdots$ \\
    M 31 & $-0.55 \pm 0.03$ & $\phantom{0}9.77 \pm 0.30$ & $38.21 \pm 0.14$ & $10.32 \pm 0.30$ & $-0.90 \pm 0.14$ & $-17.23 \pm 0.30$ & $\cdots$ \\
    M 33 & $-0.54 \pm 0.03$ & $\phantom{0}9.37 \pm 0.30$ & $38.30 \pm 0.09$ & $\phantom{0}9.91 \pm 0.30$ & $-0.82 \pm 0.09$ & $-16.82 \pm 0.30$ & $\cdots$ \\
    Milky Way & $\phantom{-}0.28 \pm 0.01$ & $10.02 \pm 0.30$ & $38.91 \pm 0.13$ & $\phantom{0}9.74 \pm 0.30$ & $-1.03 \pm 0.13$ & $-16.65 \pm 0.30$ & $\cdots$ \\
    \hline
    \multicolumn{8}{c}{Starbursts} \\ \hline
    NGC 253 & $\phantom{-}0.70 \pm 0.07$ & $\phantom{0}9.57 \pm 0.30$ & $40.12 \pm 0.07$ & $\phantom{0}8.87 \pm 0.31$ & $-0.24 \pm 0.10$ & $-15.78 \pm 0.31$ & $-16.19 \pm 0.31$ \\
    M 82 & $\phantom{-}1.02 \pm 0.07$ & $\phantom{0}9.62 \pm 0.30$ & $40.19 \pm 0.07$ & $\phantom{0}8.60 \pm 0.31$ & $-0.49 \pm 0.10$ & $-15.51 \pm 0.31$ & $-16.17 \pm 0.31$ \\
    NGC 2146 & $\phantom{-}1.15 \pm 0.17$ & $\phantom{0}9.56 \pm 0.30$ & $40.81 \pm 0.18$ & $\phantom{0}8.41 \pm 0.35$ & $\phantom{-}0.00 \pm 0.25$ & $-15.32 \pm 0.35$ & $-15.49 \pm 0.35$ \\
    NGC 1068 & $\phantom{-}1.36 \pm 0.16$ & $\phantom{0}9.42 \pm 0.30$ & $40.96 \pm 0.16$ & $\phantom{0}8.06 \pm 0.34$ & $-0.06 \pm 0.23$ & $-14.97 \pm 0.34$ & $-15.20 \pm 0.34$ \\
    Arp 299 & $\phantom{-}1.99 \pm 0.06$ & $10.14 \pm 0.30$ & $41.46 \pm 0.14$ & $\phantom{0}8.15 \pm 0.31$ & $-0.19 \pm 0.15$ & $-15.06 \pm 0.31$ & $-15.42 \pm 0.33$ \\
    Arp 220 & $\phantom{-}2.33 \pm 0.07$ & $\phantom{0}9.41 \pm 0.30$ & $42.36 \pm 0.09$ & $\phantom{0}7.08 \pm 0.31$ & $\phantom{-}0.37 \pm 0.11$ & $-13.99 \pm 0.31$ & $-13.79 \pm 0.31$ \\
    \hline
    \end{tabular}
    \caption{Measured and inferred galaxy properties; galaxies have been roughly sorted into normal star forming galaxies ($\tdep > 1$ Gyr) and starbursts ($\tdep < 1$ Gyr). Columns are as follows: (1) galaxy name; (2) star formation rate; (3) total mass mass; (4) $\gamma$-ray luminosity over the (0.1,100) GeV band; (5) depletion time $\tdep=\Mg/\dot{M}_*$; (6) calorimetry fraction from \autoref{eq:fcal}; (7) primary ionisation rate per H nucleon derived from $\tdep$ (\autoref{eq:zeta_tot}); (8) primary ionisation rate per H nucleon derived from $L_\gamma/\Mg$ for full calorimetry (\autoref{eq:zeta_gamma}).
    Star formation rate and $\gamma$-ray luminosities are taken from Table 1 of \citet{Kornecki20a}. Gas masses are from the following sources: SMC and LMC -- \citet{Jameson16a}; M31 -- \citet{Chemin09a}; M33 -- \citet{Kam17a}; Milky Way -- \citet{Kalberla09a}; all starbursts -- \citet{Liu15a}, with an extra contribution of the H~\textsc{i} mass taken from \citet{de-Blok18a} for NGC 253. Uncertainties in $\dot{M}_*$ and $L_\gamma$ are as reported in the original sources, while for gas masses, where in most cases the authors to not provide an uncertainty estimate, we adopt an uncertainty of a factor of 2 (0.3 dex). Uncertainties on the remaining quantities are determined from error propagation.
    \label{tab:galaxies}
    }
\end{table*}

The $\gamma$-ray calorimetry results shown in \autoref{tab:galaxies} are qualitatively similar to those found by previous authors \citep{Kornecki20a, Crocker21a}, which is not surprising given that our new calibration for the $\gamma$-ray emission budget only differs from past ones by a factor $<2$. We find that weakly star-forming galaxies like the Milky Way and Andromeda sit at $\approx 10\%$ of calorimetry, while starbursts such as NGC 253 and NGC 2146 sit near 100\% calorimetry. We find that Arp 220 is slightly supercalorimetric ($f_\mathrm{cal}=2.3$), but given the substantial systematic uncertainties in both its star formation rate and $\gamma$-ray luminosity (which are much larger than the statistical errors shown in the table), as well as the substantial theoretical uncertainties in quantities such as $\epsilon_p$, this result is not terribly concerning.

The ionisation results are more interesting. We find that normal star-forming galaxies have CR ionisation budgets comparable to that of the Milky Way $\zeta_{\dot{M}_*}\approx 10^{-17}-10^{-16}$ s$^{-1}$. The results for the starbursts are more interesting: while the ionisation rate budgets are certainly higher than for normal galaxies, with the exception of Arp 220 they are larger than those of the normal star-forming galaxies by only about an order of magnitude, i.e., typically $\zeta_{\dot{M}_*}\sim 10^{-16}-10^{-15}$ s$^{-1}$ rather than $\sim 10^{-17}-10^{-16}$ s$^{-1}$. The fundamental reason is that the ionisation budget scales as $1/\tdep$, and while these galaxies have depletion times shorter than those of ordinary star-forming galaxies, the depletion time for starbursts differs from that of star-forming galaxies by much less than the star formation rate per unit area. Qualitatively, if galaxies follow a \citet{Kennicutt98a}-like relation $\dot{\Sigma}_*\propto \Sigma_\mathrm{g}^{1.4}$, then the depletion time only decreases with surface density as $\Sigma_\mathrm{g}^{-0.4}$ -- thus in going from the Milky Way, at $\Sigma_\mathrm{g}\sim 10$ M$_\odot$ pc$^{-2}$, to the most extreme starbursts (such as Arp 220), $\Sigma_\mathrm{g}\sim 10^4$ M$_\odot$ pc$^{-2}$, the ionisation budget increases by only a factor of $\sim 20$. Indeed, we could deduce as much simply from \autoref{eq:zeta_tot}: even if, based on our findings for the Milky Way, we assume that additional sources of CR power not linked directly to star formation can increase the CR ionisation budget provided by star formation by a factor of several, achieving a mean ionisation rate as high as $10^{-12}$ s$^{-1}$ on galactic scales as some authors have contemplated \citep[e.g.,][]{Bisbas17a, Gonzalez-Alfonso18a} would require star formation with a depletion time $\tdep \lesssim 1$ Myr to power it. This is shorter than the depletion time of \textit{any} known galactic-scale star-forming system.

\subsection{$\gamma$-ray emission as an ionisation diagnostic}

A third implication of our calculation is that, for dense galaxies where proton calorimetry is a reasonable assumption, one can use the $\gamma$-ray luminosity per unit mass of a system as a rough diagnostic of its ionisation budget. This works particularly well for $\gamma$-ray emission measured in the (0.1,100) GeV band, since, as shown in \aref{app:differential_contributions}, in this case the energy range that gives rise to the $\gamma$-ray signal is not all that different from that which gives rise to the ionisation signal. For simplicity, since this calculation is approximate, let us consider only emission and ionisation as both being due to a single dominant mechanism. With this simplification, taking the ratio of equations \autoref{eq:Lgamma_m} and \autoref{eq:zeta_m} yields
\begin{eqnarray}
    \zeta & = & \frac{\mu_\mathrm{H} m_\mathrm{H}}{I} \left\{ \frac{\Psiip[1+\delta (\Psiie/\Psiip)]}{\Psiep [1+\delta(\Psiee/\Psiep)]}\right\}\frac{L_\gamma}{\Mg}
    \label{eq:zeta_gamma}
    \\
    & \approx & 
    1.8\times 10^{-16} \left(\frac{L_\gamma/\Mg}{10^{40}\,\mbox{erg s}^{-1}/10^9\,\mathrm{M}_\odot}\right)
    \mbox{ s}^{-1},
    \nonumber
\end{eqnarray}
where the numerical evaluation in the second line is for our fiducial values of the efficiencies, a $\gamma$-ray band pass of (0.1,100) GeV, and a background medium of H$_2$. Values for an H~\textsc{i} medium and for a (1,1000) GeV band pass can be obtained by plugging the appropriate efficiencies into the expression above, but differ only slightly in their numerical values from the case shown.

\begin{figure}
    \includegraphics[width=\columnwidth]{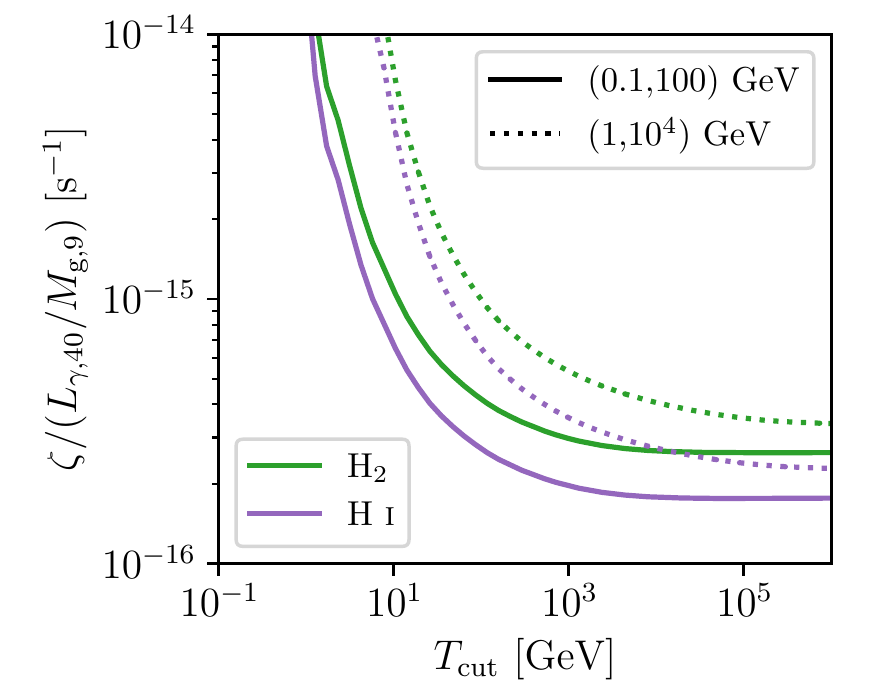}
    \caption{
    Ratio of ionisation budget $\zeta$ to $\gamma$-ray luminosity per unit gas mass, $L_\gamma/\Mg$, as a function of maximum CR injection energy $T_{\rm cut}$. We normalise the $\gamma$-ray luminosity per unit gas mass by expressing $\gamma$-ray luminosity as $L_{\gamma,40} = L_\gamma/10^{40}$ erg s$^{-1}$, and gas mass as $M_{\mathrm{g},9} = \Mg/10^9$ M$\odot$. Sold lines show results for a (0.1,100) GeV $\gamma$-ray band pass, dotted lines for a (1,$10^4$) GeV band pass. Green shows results for a background of pure H$_2$, purple for a background of pure H~\textsc{i}. All calculations use an electron-to-proton luminosity ratio $\delta=0.1$.
    \label{fig:Psi_ratio}
    }
\end{figure}

This result is a useful complement to our estimate of the ionisation rate $\zeta_{\dot{M}_*}$ from the star formation rate, because that result depends on details of star formation and ISM physics such as the number of supernovae per unit mass of stars formed and the CR acceleration efficiency. By contrast, these factors all cancel in \autoref{eq:zeta_gamma}: the only assumptions that enter this equation are that the galaxy emitting the $\gamma$-rays is calorimetric, and that ionisation and $\gamma$-ray emission are both driven mainly by mechanisms with high cutoff energies $T_{\rm cut}$ such as SNe. The ionisation budget could be higher if either of these assumptions fail -- if for example the observed value of $L_\gamma$ does not reflect the true $\gamma$-ray energy budget because some CRs escape the galaxy, or if there are significant CR sources with $T_{\rm cut}$ low enough that they do not produce $\gamma$-rays but still produce ionisation. We quantify the latter possibility by plotting the ratio $\zeta/(L_\gamma/\Mg)$ as a function of $T_{\rm cut}$ in \autoref{fig:Psi_ratio}. Clearly the $\gamma$-ray diagnostic of ionisation fails completely for $T_{\rm cut} \lesssim 1$ GeV (for the \textit{Fermi}-like band pass; $\lesssim 10$ GeV for the CTA-like one), since in this case essentially no $\gamma$-rays within the band pass are produced. However, the plot also shows that \autoref{eq:zeta_gamma} is reasonably reliable as long as ionisation is not dominated by sources with $T_\mathrm{cut} \lesssim 10$ GeV. Quantitatively, for the (0.1,100) GeV band pass, the ratio $\zeta/(L_\gamma/\Mg)$ varies by less than a factor of 4 as $T_{\rm cut}$ goes from $10$ GeV to infinity (and by less than a factor of 2 for $T_{\rm cut}=40$ GeV to infinity); it also differs by only a factor $1.7$ for H~\textsc{i} versus H$_2$ backgrounds.

We derive alternative estimates of the ionisation budget for starburst galaxies from \autoref{eq:zeta_gamma} and list the results as $\zeta_{L_\gamma/\Mg}$ in \autoref{tab:galaxies}; we do so only for the starburst galaxies where full calorimetry is a reasonable assumption. Qualitatively, these estimates are similar to those derived from the star formation rate, which is not surprising since \autoref{eq:zeta_gamma} is derived under the assumption of full calorimetry, and we find that starbursts are close to this limit. The point of this exercise is simply that it eliminates most of the systematics listed above, e.g., unknown CR acceleration efficiencies, number of SN production per unit star formation, contributions from non-SN sources, etc. The \textit{only} assumptions that enter estimates of the ionisation budget from $\gamma$-ray luminosities and gas masses are that starburst galaxies are calorimetric and that the CR injection spectrum follows the usual powerlaw in momentum, with a cutoff energy $\gtrsim 10$ GeV. 

Our results therefore reinforce the conclusion that the primary ionisation rates in starburst galaxies are elevated compared to those in normal galaxies, but not by as much as some proposals in the literature suggest \citep[e.g.,][]{Papadopoulos10a, Meijerink11a, Bisbas15a, Bisbas17a, Papadopoulos18a}. For moderate starbursts such as NGC 253 or M82, the enhancement compared to the Milky Way is roughly an order of magnitude, while for the most extreme starbursts such as Arp 220 it is at most $\sim 3$ dex. The only way to escape this conclusion would be to posit that ionisation in these galaxies is driven mainly by sources that produce CRs with low maximum energies (or more generically with spectra that are not powerlaws in momentum with index $q\sim 2 - 2.5$), such that they produce ionisation but no $\gamma$-ray emission.

\section{Conclusions}
\label{sec:conclusions}

In this paper we investigate the budget for cosmic rays (CRs) accelerated by star formation to drive diffuse $\gamma$-ray emission and ionisation in galaxies. We do so using a particle-by-particle approach, whereby we compute the maximum total number of ionisations and the total emitted $\gamma$-ray energy that CR protons and electrons of a specified initial energy can produce. Integrating these production rates over the spectral distribution with which CRs are injected, and normalising by the total CR power provided by different forms of star formation feedback, then gives the maximum rates of $\gamma$-ray production and ionisation that a given star formation rate is capable of driving. 

A principal result of our calculations is that the $\gamma$-ray emission and ionisation budgets are
\begin{eqnarray}
    L_\gamma & = & 4\times 10^{39}\left(\frac{\dot{M}_*}{\mathrm{M}_\odot\mbox{ yr}^{-1}}\right)\mbox{ erg s}^{-1} \\
    \zeta & = & 1\times 10^{-16} \left(\frac{\tdep}{\mbox{Gyr}}\right)^{-1}\mbox{ s}^{-1},
\end{eqnarray}
where $\zeta$ is the primary ionisation rate per H nucleon, $L_\gamma$ is the $\gamma$-ray luminosity, $\dot{M}_*$ is the galactic star formation rate, and $\tdep$ is the galactic gas depletion time -- see \autoref{eq:zeta_tot}, \autoref{eq:Lgamma_tot}, and \autoref{tab:coefficients} for precise numbers as a function of ISM chemical state and $\gamma$-ray bandpass, and for a decomposition of the budgets into different CR acceleration mechanisms. Our value of $L_\gamma$, while improved compared to earlier calculations due to more realistic treatments of supernovae, a more extended set of microphysical processes included, and updated cross-section data, differs from earlier results by less than a factor of 2, and leads to qualitatively similar conclusions when used to analyse observations: normal star-forming galaxies such as the Milky Way typically radiate only $\approx 10\%$ of their available $\gamma$-ray budget, indicating that many CR protons escape, while starbursts are calorimetric or close to it.

By contrast, our calculation of the ionisation budget is novel and leads to more interesting conclusions. We find that the available ionisation budget is too small by a factor of a few to produce mean ionisation rates as high as those measured in Milky Way molecular clouds. This indicates either that molecular material has an elevated ionisation rate compared to the mean of neutral gas in the Galaxy (plausible, since stellar winds and protostellar jets make a significant contribution to the ionisation budget, and this contribution is likely concentrated in molecular clouds), or that there are additional contributions to CR ionisation by sources not directly linked to recent star formation, for example type Ia SNe or second-order Fermi acceleration, though we disfavour the latter possibility on energetic grounds. A corollary of this analysis is that the Galaxy is consuming most of its available CR ionisation budget. Unlike for $\gamma$-ray-producing CRs (those with kinetic energies $\approx 1 - 10^3$ GeV), where 90\% of the CR energy escapes into the halo, most of the energy carried by the trans-relativistic CRs that dominate the ionisation budget (those with kinetic energies $\lesssim m_p c^2$) must be dissipated within the Galaxy. The conclusion is confirmed by the fact that the observed spectral shape for low-energy ($\lesssim 100$ MeV) protons and electrons in the local ISM matches that expected for injection of CRs into a thick target.

As applied to external galaxies, our calculation of the budget implies that the ionisation rates in the bulk of starburst galaxy interstellar media can be elevated only mildly compared to that in the Milky Way. Ionisation rates in moderate starbursts such as NGC 253 or M82 are likely a factor of $\approx 10$ above that in the Milky Way, while those in the most extreme starbursts such as Arp 220 can reach a few hundred times Milky Way values. The fundamental factor driving these results is that the Milky Way is already near its ionisation budget, and the ionisation budget scales only with the gas depletion time. While starbursts often have star formation rates per unit area or per unit volume larger than that of the Milky Way by factor of $>1000$, their depletion times differ from the Milky Way depletion time by a much smaller factor.

Finally, we point out that, in galaxies that can reasonably be approximated as reaching full proton calorimetry, the $\gamma$-ray luminosity per unit gas mass provides a direct estimate of the ionisation rate (see \autoref{eq:zeta_gamma}). This estimator is valid as long as the dominant sources of CRs in a galaxy produce a powerlaw momentum distribution similar to that expected for shocks, with a cutoff energy $\gtrsim 10$ GeV, and has the advantage that it is essentially independent of ISM or star formation physics; it depends only on microphysical cross sections. Use of this alternative estimator confirms our results for the modest ionisation rate enhancements in starbursts and offers a new method to constrain astrochemical conditions in galaxies where more direct estimates of the CR-driven ionisation rate are unavailable. 

\section*{Data availability}

The \textsc{criptic} CR simulation software used for the numerical simulations in this paper is freely available from \url{https://bitbucket.org/krumholz/criptic/src/master/}. The \textsc{criptic} input files and analysis scripts that generate all the quantitative results and plots in the paper, along with summary files from the \textsc{criptic} simulations, are available from \url{https://bitbucket.org/krumholz/kco22}. The full \textsc{criptic} outputs are not included in the repository due to their size, but are available upon reasonable request to MRK. The \textsc{slug} software used for the star formation budget calculations is available from \url{https://bitbucket.org/krumholz/slug2/src/master/}.

\section*{Acknowledgements}

We acknowledge Jessica Ross and Melanie Rosevear for assistance in the early stages of this project. MRK and RMC acknowledge support from the Australian Research Council through its \textit{Discovery Projects} funding scheme, award DP190101258, and from resources and services from the National Computational
Infrastructure (NCI), award jh2, which is supported by the Australian Government. SSRO acknowledges support from NASA ATP 80NSSC20K0507 and the Stromlo Distinguished Visitor program.  



\bibliographystyle{mnras}
\bibliography{refs,newrefs} 




\begin{appendix}

\section{Effects of varying synchrotron and inverse Compton losses}
\label{app:fsyncIC}

Throughout the main body of the paper, we present results for $f_\mathrm{sync} = f_\mathrm{IC} = 10^{-7}$ (c.f.~\autoref{eq:fsync}), values that we conclude are typical of both normal and starburst galaxies. To explore the sensitivity of our results to this assumption, we repeat the simulations presented in \autoref{ssec:criptic_numerical} with two alternative values, $f_\mathrm{sync} = f_\mathrm{IC} = 10^{-7.5}$ and $10^{-6.5}$; the corresponding energy densities in the magnetic field and radiation field, given our density $n_\mathrm{H} = 10^3$ cm$^{-3}$, are 1.2 keV cm$^{-3}$ and 120 eV cm$^{-3}$, respectively. For these cases we set up our grid of \textsc{criptic} simulations exactly as described in the main paper, simply with different values for the magnetic field strength and radiation field dilution factor, tuned to produce the desired values of $f_\mathrm{sync}$ and $f_\mathrm{IC}$.\footnote{The only other difference is that for electron energies above $10^5$ GeV in the $f_\mathrm{sync} = f_\mathrm{IC} = 10^{-7.5}$ case we reduce the secondary sampling factor $f_\mathrm{sec}$ from 0.2 to 0.1. This change does not impact the physics being simulated, and is made solely for computational convenience, to avoid having to follow a very large number of secondary sample packets in a case where the catastrophic loss mechanism of bremsstrahlung is dominant compared to the mostly continuous inverse Compton and synchrotron mechanisms; see \citet{Krumholz22a} for details in the meaning of $f_\mathrm{sec}$ in \textsc{criptic} simulations.} We then compute the spectrally-averaged ionisation and $\gamma$-ray production efficiencies for these cases exactly as in \autoref{ssec:spectral_average}.

\begin{figure}
    \includegraphics[width=\columnwidth]{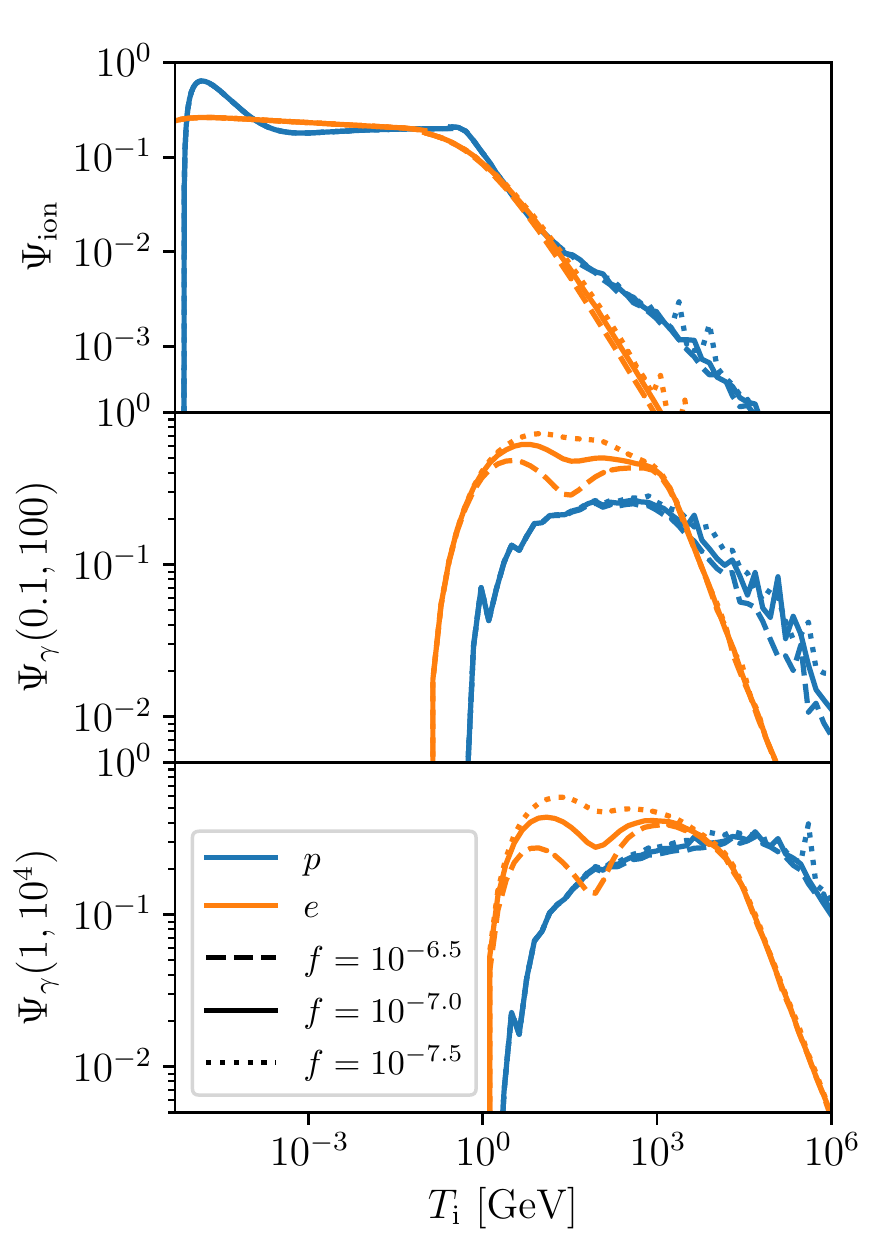}
    \caption{
    Ionisation and $\gamma$-ray production efficiencies $\Psi$ as a function of initial CR kinetic energy $T_i$, computed using different value of $f_\mathrm{sync/IC}$ in an H$_2$ background. The top panel shows $\Psi_\mathrm{ion}$ and the middle and bottom panels show $\Psi_\gamma$ computed over the $(0.1,100)$ and $(1,10^4)$ GeV band passes, respectively. Blue lines show protons and orange lines show electrons. Solid lines show our fiducial case $f_\mathrm{sync} = f_\mathrm{IC} = 10^{-7}$, and are identical to the lines shown in \autoref{fig:NionLp} and \autoref{fig:NionLe} in the main text; dashed and dotted lines show $f_\mathrm{sync} = f_\mathrm{IC} = 10^{-6.5}$ and $10^{-7.5}$, respectively.
    \label{fig:Psi_f_comp}
    }
\end{figure}

We compare ionisation and $\gamma$-ray production efficiencies for our fiducial case and for the two alternative cases in \autoref{fig:Psi_f_comp}, which is analogous to \autoref{fig:NionLp} and \autoref{fig:NionLe} in that it shows $\Psi_\mathrm{ion}$ and $\Psi_\gamma$ as a function of CR energy; the figure shows only the case for an H$_2$ background medium, but the results for H~\textsc{i} are qualitatively identical. The primary conclusion to be drawn from the figure is that changing $f_\mathrm{sync}$ and $f_\mathrm{IC}$ by half of dex on either side of our fiducial value induces completely negligible changes in the ionisation efficiency for either protons or electrons. This is not surprising given that ionisations occur primarily at low energies where synchrotron and inverse Compton scattering are unimportant; changing $f_\mathrm{sync}$ and $f_\mathrm{IC}$ may change the amount of time that an individual CR electron takes to lose enough energy for ionisation losses to become dominant, but in the limit of a thick target where no CRs escape, ultimately they do not change the amount of energy that is available to go into ionisation. The largest effect of varying $f_\mathrm{sync}$ and $f_\mathrm{IC}$ is to change the $\gamma$-ray production efficiency for $\sim 10$ GeV electrons. For these particles, factor of ten variations in $f_\mathrm{sync}$ and $f_\mathrm{IC}$ induce factor of $\sim 3$ variations in $\gamma$-ray production. A smaller but related effect is also visible for $\gtrsim 10$ TeV protons producing $0.1 - 100$ GeV $\gamma$-rays, where the efficiency depends on $f_\mathrm{sync}$ and $f_\mathrm{IC}$ because secondary electrons make a subdominant but non-negligible contribution to the emission. However, here order of magnitude changes in $f_\mathrm{sync}$ and $f_\mathrm{IC}$ only produce $\approx 10\%$ changes in $\Psi_\gamma$. Overall, the conclusion to be drawn from \autoref{fig:Psi_f_comp} is that plausible variations in $f_\mathrm{sync}$ and $f_\mathrm{IC}$ produce relatively small changes in production efficiencies.

\begin{table}
    \begin{center}
   \begin{tabular}{lccc}
    \hline\hline
    Quantity & $f = 10^{-7.5}$ & $f=10^{-7}$ & $f=10^{-6.5}$ \\
    \hline
         \hline
         \multicolumn{4}{c}{H$_2$ background}
         \\ \hline
         $\Psiipavg$ & $0.058$ & $0.058$ & $0.058$ \\
         $\Psiieavg$ & $0.185$ & $0.185$ & $0.185$ \\
         $\Psiepavg(0.1,100)$ & $0.143$ & $0.139$ & $0.134$ \\
         $\Psiepavg(1,10^4)$ & $0.116$ & $0.111$ & $0.107$ \\
         $\Psieeavg(0.1,100)$ & $0.098$ & $0.086$ & $0.071$ \\
         $\Psieeavg(1,10^4)$ & $0.055$ & $0.043$ & $0.030$ \\
         \hline
         \multicolumn{4}{c}{H~\textsc{i} background}
         \\ \hline
         $\Psiipavg$ & $0.030$ & $0.030$ & $0.030$ \\
         $\Psiieavg$ & $0.155$ & $0.155$ & $0.155$ \\
         $\Psiepavg(0.1,100)$ & $0.143$ & $0.140$ & $0.136$ \\
         $\Psiepavg(1,10^4)$ & $0.115$ & $0.111$ & $0.108$ \\
         $\Psieeavg(0.1,100)$ & $0.099$ & $0.087$ & $0.071$ \\
         $\Psieeavg(1,10^4)$ & $0.056$ & $0.043$ & $0.030$ \\
    \hline\hline
    \end{tabular}
    \end{center}
    \caption{
    Spectrally-averaged ionisation and $\gamma$-ray emission efficiencies computed using alternative values of $f_\mathrm{sync} = f_\mathrm{IC} = f$. For the latter, $\Psiepavg(0.1,100)$ and $\Psiepavg(1,10^4)$ refer to efficiencies integrated oer the $(0.1,100)$ and $(1,10^4)$ GeV band passes, and similarly for $\Psieeavg$.
    \label{tab:fsyncIC}
    }
\end{table}

To quantify this conclusion we repeat our calculation of the spectrally-averaged efficiencies $\Psiipavg$, $\Psiepavg$, $\Psiepavg$, and $\Psieeavg$ using our alternative values of $f_\mathrm{sync}$ and $f_\mathrm{IC}$, for our fiducial choices $T_\mathrm{cut} = 10^6$ TeV and $q=2.25$. We show the results of this calculation in \autoref{tab:fsyncIC}. The table shows that, when we average over the full injected CR spectrum, factor of $10$ changes in $f_\mathrm{sync}$ and $f_\mathrm{IC}$ make $<1\%$ differences in the ionisation efficiency, $\approx 10\%$ differences in the $\gamma$-ray production efficiency for electrons, and $\approx 1\%$ differences in the $\gamma$-ray production efficiency for protons.

\section{\add{Ionisation and $\gamma$-ray production efficiencies as a function of CR energy}}
\label{app:differential_contributions}

\begin{figure}
    \includegraphics[width=\columnwidth]{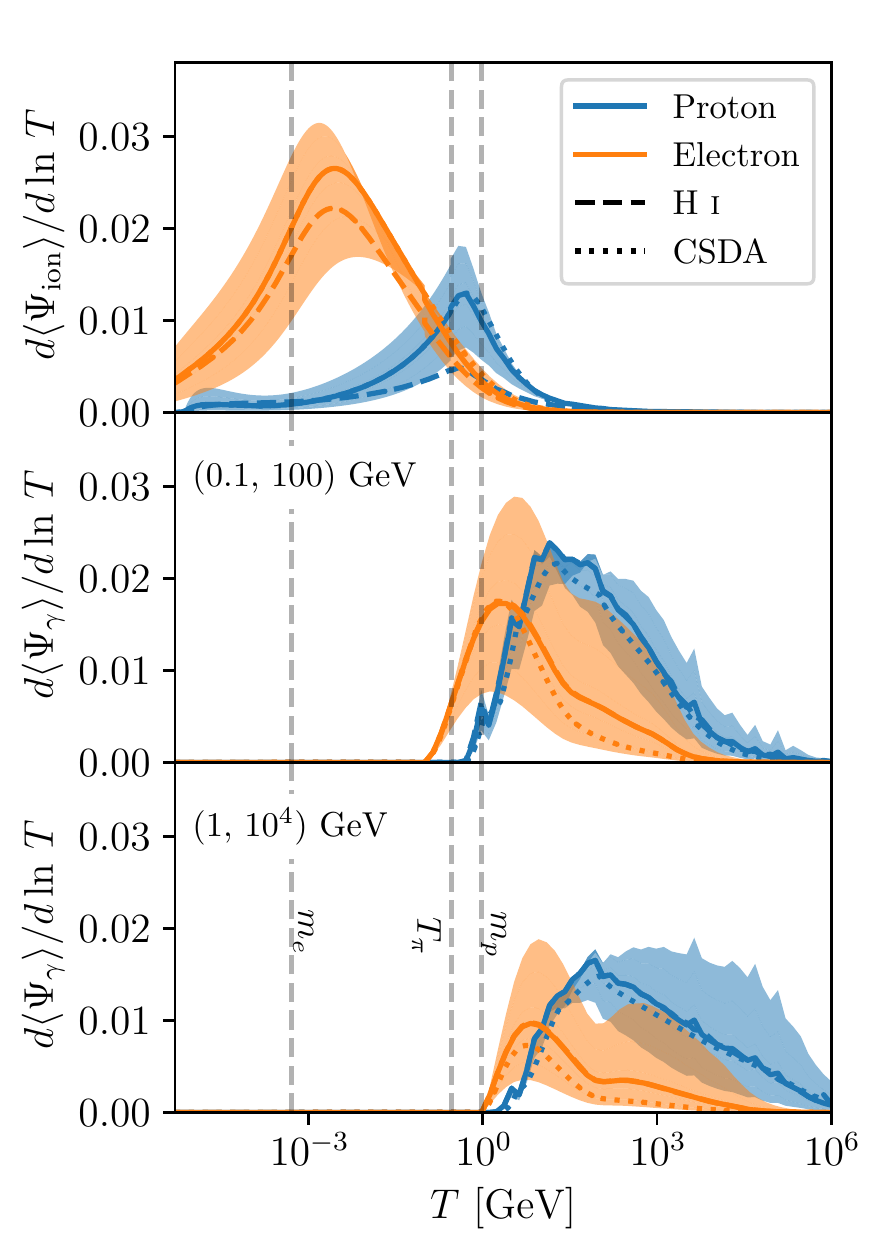}
    \caption{
    \label{fig:dPsidlnT}
    The marginal contribution of CR protons (blue) and electrons (orange) with initial kinetic energies $T$ to ionisation, $d\langle\Psi_\mathrm{ion}\rangle/d\ln T$ (top panel), and $\gamma$-ray emission, $d\langle\Psi_\gamma\rangle/d\ln T$; solid lines show the full numerical result obtained using \textsc{criptic} for propagation through a medium where all hydrogen is in the form of H$_2$, dashed lines show the result for a medium of all H~\textsc{i}, and dotted lines show the results for H$_2$ computed using the CSDA. For $d\Psi_\gamma/d\ln T$, we show the $\gamma$-ray luminosity integrated from $0.1$ to 100 GeV in the middle panel and from 1 to $10^4$ GeV in the bottom panel. In all panels, the central solid line marks the result for a CR spectral index $q=2.25$, and the shaded region indicates the range for $q\in(2.1, 2.4)$. The vertical dashed lines mark, from left to right, the electron rest mass, pion production threshold for CR protons, and proton rest mass. 
    }
\end{figure}

\add{In the main text we compute the spectrally-averaged ionisation and $\gamma$-ray production efficiencies from \autoref{eq:Psiipavg} and \autoref{eq:Psiepavg}. While these are the primary quantities of astrophysical interest, for the purposes of interpreting the results it is helpful to examine the differential contribution of CRs with different initial energies to ionisation and $\gamma$-ray emission. To do so, in \autoref{fig:dPsidlnT} we plot} the integrands of the integrals in \autoref{eq:Psiipavg} and \autoref{eq:Psiepavg}, and their electron equivalents, as a function of CR kinetic energy $T$. For plotting convenience we change the integration variable from $x$ or $y$ to $\ln T$, i.e., we plot the contribution to $\Psi$ per unit $\ln T$. Specifically, for proton-induced ionisations we plot
\begin{equation}
    \frac{d\Psiipavg}{d\ln T} = \phi_p \Psiip x^{-q} \left(\frac{x}{d\ln T/d\ln p}\right),
\end{equation}
where the factor in parentheses is what is required to convert from an integral with respect to $x$, as in \autoref{eq:Psiipavg}, to an integral with respect to $\ln T$. The expressions for electron-driven ionisations, and for proton- and electron-driven $\gamma$-ray production, are analogous. We plot these quantities for a fiducial spectral index $q = 2.25$, and show the results for a range from $q = 2.1$ to $2.4$.

From this figure, we see that ionisations in either an H$_2$- or H~\textsc{i}-dominated medium are mostly driven by trans-relativistic CRs, with energies relatively close to the particle rest energy. This is simply a consequence of this being the locus that carries most of the CR energy: for a momentum distribution $d\dot{n}/dp \propto p^{-q}$, the corresponding energy distribution is $d\dot{n}/dT \propto T^{-(q+1)/2}$ in the non-relativistic regime and $d\dot{n}/dT\propto T^{-q}$ in the ultra-relativistic regime. For $q$ in the plausible range of $2.1$ to $2.4$, this means that the index of the energy distribution $d\dot{n}/dT$ is shallower than $2$ at low energies and steeper than $2$ at higher energies, indicating that the bulk of the energy must reside in the trans-relativistic regime that marks the transition between shallow and steep energy powerlaws. \add{A corollary of this analysis is that, as long as we choose a lower energy cutoff for the CR injection distribution that is well into the non-relativistic regime, our results are insensitive to the exact value we choose, since all of the energy resides near the trans-relativistic peak.} In the case of protons, the peak in the trans-relativistic regime is further sharpened by pion losses, which suppress the contribution from higher energy CRs by siphoning the available energy into another loss channel. We also see that, for ionisation, the CSDA is extremely accurate. The only visible differences between the full numerical and CSDA results are for protons at energies $\gtrsim 1$ GeV, where secondaries become significant, but even these differences are at the $\sim 10\%$ level.

For $\gamma$-ray production, to first order we see that the energy range that contributes roughly matches the band pass of the observations. However, the CR energy range that contributes is somewhat narrower than this, precisely because the total available energy is falling off as one moves to higher energy. The steepness of this falloff depends on $q$, particularly in the higher-energy band pass that resembles the CTA sensitivity range. Nonetheless, an important conclusion to draw from \autoref{fig:dPsidlnT} is that, particularly for protons (which are expected to dominate simply because they dominate the overall energy budget), the range of CR energies that drives ionisation is not that far from the range that drives $\gamma$-ray emission. We also see that the CSDA is quite accurate for protons, but somewhat underestimates electron emission. This underestimate, however, is still within the plausible range corresponding to variations in the CR spectral index. Finally, the results for a background medium or H~\textsc{i} or H$_2$ are so similar that the lines are essentially indistinguishable in \autoref{fig:dPsidlnT}.

\section{Tabulated efficiencies}
\label{app:tabulation}

\add{In \autoref{tab:Psi}, for reader convenience we tabulate our computed spectrally-averaged efficiencies for ionisation and $\gamma$-ray production for protons and electrons as a function of $T_\mathrm{cut}$ and $q$.}

\begin{table*}
    \centering
    \begin{tabular}{llcccccc}
    \hline\hline
         $q$ & $T_\mathrm{cut}$ & $\Psiipavg$ & $\Psiieavg$ & \multicolumn{2}{c}{$\Psiepavg$} & \multicolumn{2}{c}{$\Psieeavg$} \\
         & [GeV] & & & (0.1,100) GeV & (1,$10^4$) GeV & (0.1,100) GeV & (1,$10^4$) GeV \\
         \hline
         \multicolumn{8}{c}{H$_2$ background}
         \\ \hline
         2.25 & $10^6$ & 0.058 & 0.185 & 0.139 & 0.111 & 0.086 & 0.043 \\
         2.10 & $10^6$ & 0.031 & 0.125 & 0.152 & 0.167 & 0.176 & 0.114 \\
         2.40 & $10^6$ & 0.090 & 0.213 & 0.109 & 0.067 & 0.033 & 0.013 \\
         2.25 & $10^{-1}$ & 0.203 & 0.224 & 0.000 & 0.000 & 0.000 & 0.000 \\
         2.25 & $10^{1}$ & 0.122 & 0.203 & 0.058 & 0.008 & 0.054 & 0.013 \\
         2.25 & $10^{3}$ & 0.069 & 0.190 & 0.138 & 0.079 & 0.084 & 0.037 \\
         2.25 & $10^{5}$ & 0.060 & 0.186 & 0.142 & 0.108 & 0.087 & 0.043 \\
         \hline
         \multicolumn{8}{c}{H~\textsc{i} background}
         \\ \hline
         2.25 & $10^6$ & 0.030 & 0.155 & 0.140 & 0.111 & 0.087 & 0.043 \\
         2.10 & $10^6$ & 0.015 & 0.104 & 0.152 & 0.168 & 0.178 & 0.115 \\
         2.40 & $10^6$ & 0.051 & 0.180 & 0.110 & 0.067 & 0.033 & 0.013 \\
         2.25 & $10^{-1}$ & 0.149 & 0.189 & 0.000 & 0.000 & 0.000 & 0.000 \\
         2.25 & $10^{1}$ & 0.063 & 0.170 & 0.060 & 0.008 & 0.055 & 0.013 \\
         2.25 & $10^{3}$ & 0.036 & 0.159 & 0.139 & 0.079 & 0.085 & 0.037 \\
         2.25 & $10^{5}$ & 0.031 & 0.156 & 0.143 & 0.108 & 0.087 & 0.043 \\
    \hline\hline
    \end{tabular}
    \caption{Values of mean ionisation efficiency $\langle\Psi_{\rm ion}\rangle$ and $\gamma$-ray production efficiency $\langle\Psi_\gamma\rangle$ for protons and electrons for sample values of the parameters $q$ and $T_{\rm cut}$ describing the injection spectrum; $\langle\Psi_\gamma\rangle$ is shown for band passes of both (0.1,100) and (1,$10^4$) GeV. The top block of values is for a background medium of pure H$_2$, the bottom block for pure H~\textsc{i}.}
    \label{tab:Psi}
\end{table*}

\end{appendix}


\bsp	
\label{lastpage}
\end{document}